\documentclass[a4paper,12pt]{article}
\pdfoutput=1
\usepackage{epsfig}
\usepackage{amssymb}
\usepackage{amsfonts}
\usepackage{amsmath}
\usepackage{euscript}
\usepackage{verbatim}
\usepackage{latexsym}
\usepackage{graphicx}
\usepackage{caption}
\usepackage{float}
\usepackage{subcaption}

\usepackage{listings}

\newif\ifdtup

\catcode`\@=11

\@addtoreset{equation}{section}

\def\@normalsize{\@setsize\normalsize{15pt}\xiipt\@xiipt
\abovedisplayskip 14pt plus3pt minus3pt%
\belowdisplayskip \abovedisplayskip
\abovedisplayshortskip \z@ plus3pt%
\belowdisplayshortskip 7pt plus3.5pt minus0pt}

\def\small{\@setsize\small{13.6pt}\xipt\@xipt
\abovedisplayskip 13pt plus3pt minus3pt%
\belowdisplayskip \abovedisplayskip
\abovedisplayshortskip \z@ plus3pt%
\belowdisplayshortskip 7pt plus3.5pt minus0pt
\def\@listi{\parsep 4.5pt plus 2pt minus 1pt
     \itemsep \parsep
     \topsep 9pt plus 3pt minus 3pt}}

\relax

\catcode`@=12

\catcode`\@=11

\def\section{\@startsection{section}{1}{\z@}{3.5ex plus 1ex minus
   .2ex}{2.3ex plus .2ex}{\large\bf}}

\def\SymBoxes#1#2#3#4{\newdimen\un@t \un@t#3%
\raisebox{#1}{\rule{#2\un@t}{#4}\hskip-#2\un@t% lower horizontal
\@tempdimb\un@t \advance\@tempdimb by-#4\@tempcntb#2\relax%
\@whilenum{\@tempcntb>0}\do{%                         % #2 vertical lines
\rule{#4}{\un@t}\hskip\@tempdimb \advance\@tempcntb by\m@ne}%
\hskip-#2\un@t \rule[\un@t]{#2\un@t}{#4}%
\rule[\un@t]{#4}{#4}\hskip-#4%             % upper horizontal line
\rule{#4}{\un@t}}\hskip-#4}                % rightest vertical line
\begin{document}
%\begin{letter}{~}

%%%%%%Define some new commands and  macros
\newcommand{\beq}{\begin{equation}}
\newcommand{\eeq}{\end{equation}}
\newcommand{\bea}{\begin{eqnarray}}
\newcommand{\eea}{\end{eqnarray}}
\newcommand{\beas}{\begin{eqnarray*}}
\newcommand{\eeas}{\end{eqnarray*}}
\newcommand{\defi}{\stackrel{\rm def}{=}}
\newcommand{\non}{\nonumber}
\newcommand{\bquo}{\begin{quote}}
\newcommand{\enqu}{\end{quote}}
%%%%%%%%%%%%%%%%
\renewcommand{\(}{\begin{equation}}
\renewcommand{\)}{\end{equation}}
%%%%%%%%%%%%%%%%%%%%%%%%%%%%%%%%%% definitions
\def \eqn#1#2{\begin{equation}#2\label{#1}\end{equation}}

\def\e{\epsilon}
\def\IZ{{\mathbb Z}}
\def\IR{{\mathbb R}}
\def\IC{{\mathbb C}}
\def\IQ{{\mathbb Q}}
\def\IH{{\mathbb H}}
\def\de{\partial}
\def\Tr{ \hbox{\rm Tr}}
\def\H{ \hbox{\rm H}}
\def\HE{ \hbox{$\rm H^{even}$}}
\def\HO{ \hbox{$\rm H^{odd}$}}
\def\K{ \hbox{\rm K}}
\def\Im{ \hbox{\rm Im}}
\def\Ker{ \hbox{\rm Ker}}
\def\const{\hbox {\rm const.}}
\def\o{\over}
\def\im{\hbox{\rm Im}}
\def\re{\hbox{\rm Re}}
\def\bra{\langle}\def\ket{\rangle}
\def\Arg{\hbox {\rm Arg}}
\def\Re{\hbox {\rm Re}}
\def\Im{\hbox {\rm Im}}
\def\exo{\hbox {\rm exp}}
\def\diag{\hbox{\rm diag}}
\def\longvert{{\rule[-2mm]{0.1mm}{7mm}}\,}
\def\a{\alpha}
\def\dag{{}^{\dagger}}
\def\tq{{\widetilde q}}
\def\p{{}^{\prime}}
\def\W{W}
\def\N{{\cal N}}
\def\hsp{,\hspace{.7cm}}

\def\br{\nonumber}
\def\IZ{{\mathbb Z}}
\def\IR{{\mathbb R}}
\def\IC{{\mathbb C}}
\def\IQ{{\mathbb Q}}
\def\IP{{\mathbb P}}
\def \eqn#1#2{\begin{equation}#2\label{#1}\end{equation}}

\newcommand{\C}{\ensuremath{\mathbb C}}
\newcommand{\Z}{\ensuremath{\mathbb Z}}
\newcommand{\R}{\ensuremath{\mathbb R}}
\newcommand{\rp}{\ensuremath{\mathbb {RP}}}
\newcommand{\cp}{\ensuremath{\mathbb {CP}}}
\newcommand{\vac}{\ensuremath{|0\rangle}}
\newcommand{\vact}{\ensuremath{|00\rangle}                    }
\newcommand{\oc}{\ensuremath{\overline{c}}}
\newcommand{\psizero}{\psi_{0}}
\newcommand{\phizero}{\phi_{0}}
\newcommand{\hzero}{h_{0}}
\newcommand{\psiin}{\psi_{\rh}}
\newcommand{\phiin}{\phi_{\rh}}
\newcommand{\hin}{h_{\rh}}
\newcommand{\rh}{r_{h}}
\newcommand{\rb}{r_{S}}
\newcommand{\psibnd}{\psi_{0}^{b}}
\newcommand{\psibndp}{\psi^{}_{1}^{b}}
\newcommand{\phibnd}{\phi_{0}^{b}}
\newcommand{\phibndp}{\phi^{}_{1}^{b}}
\newcommand{\gbnd}{g_{0}^{b}}
\newcommand{\hbnd}{h_{0}^{b}}
\newcommand{\zh}{z_{h}}
\newcommand{\zb}{z_{S}}
\newcommand{\man}{\mathcal{M}}
\newcommand{\hbr}{\bar{h}}
\newcommand{\tbr}{\bar{t}}

\begin{titlepage}
\begin{flushright}
CHEP XXXXX
%ULB-TH/09-10\\
%hep-th/yymmnnn\\
\end{flushright}
\bigskip
\def\thefootnote{\fnsymbol{footnote}}

\begin{center}
{\bf {\Large Hints of Gravitational Ergodicity:} \\
\vspace{0.2cm}
Berry's Ensemble and the Universality of the Semi-Classical Page Curve
}

%BERRY'S ENSEMBLE AND THE UNIVERASLITY OF THE PAGE CURVE

\end{center}

\bigskip
\begin{center}
Chethan KRISHNAN$^a$\footnote{\texttt{chethan.krishnan@gmail.com}},\ \  Vyshnav MOHAN$^a$\footnote{\texttt{vyshnav.vijay.mohan@gmail.com}}
\vspace{0.1in}

\end{center}

\renewcommand{\thefootnote}{\arabic{footnote}}
\vspace{-0.3cm}

\begin{center}

$^a$ {Center for High Energy Physics,\\
Indian Institute of Science, Bangalore 560012, India}\\

\end{center}
\vspace{-0.3cm}
\noindent
\begin{center} {\bf Abstract} \end{center} 

%Many of the relevant questions can be de-mystified by considering a hard sphere gas leaking slowly from a small box into a bigger box.

%Our results make it very explicit that an ensemble-averaged semi-classical Page curve is {\em not} a unique feature of gravity, and that the underlying theory need {\em not} be an ensemble.

Recent developments on black holes have shown that a unitarity-compatible Page curve can be obtained from an ensemble-averaged semi-classical approximation. In this paper, we emphasize (1) that this peculiar manifestation of unitarity is not specific to black holes, and (2) that it can emerge from a single realization of an underlying unitary theory. To make things explicit, we consider a hard sphere gas leaking slowly from a small box into a bigger box. This is a quantum chaotic system in which we expect to see the Page curve in the full unitary description, while semi-classically, eigenstates are expected to behave as though they live in Berry's ensemble. We reproduce the unitarity-compatible Page curve of this system, semi-classically. The computation has structural parallels to replica wormholes, relies crucially on ensemble averaging at each epoch, and reveals the interplay between the multiple time-scales in the problem. Working with the ensemble averaged $state$ rather than the entanglement entropy, we can also engineer an information ``paradox". Our system provides a concrete example in which the ensemble underlying the semi-classical Page curve is an ergodic proxy for a time average, and not an explicit average over many theories. The questions we address here are logically independent of the existence of horizons, so we expect that semi-classical gravity should also be viewed in a similar light.

\vspace{1.6 cm}
\vfill

\end{titlepage}

\setcounter{footnote}{0}

%%%%%%%%%%%%%%%%%%%%%%%%%%%%%%%%%%%%%%%%%%%%%%%%%%%%%%%%%%%%%%%%%%%%%%%%%%%%%%%%%%%%%%%%%%%%%%
%%%%%%%%%%%%%%%%%%%%%%%%%%%%%%%%%%%%%%%%%%%%%%%%%%%%%%%%%%%%%%%%%%%%%%%%%%%%%%%%%%%%%%%%%%%%%%
\section{Introduction}

Recent developments \cite{Penington,Almheiri} on the information paradox \cite{Hawking, Page, Mathur, AMPS} have revealed that one can reproduce the Page curve for Hawking radiation from semi-classical gravity. This can be viewed as surprising for a couple of reasons: 

\begin{itemize}
\item Firstly, it reveals that {\em understanding the fine-grained  entropy (or at least its qualitative Page evolution) does not require us to know the  microstate/density matrix in the full UV-complete theory; a knowledge of the semi-classical description is enough}. While this fact may seem superficially surprising, it should be emphasized that there is no contradiction here. Entanglement entropy is just one number, and the full density matrix is a (possibly infinite dimensional) matrix. So the latter contains a vastly larger amount of information, which is in principle not required for extracting the fine-grained entropy. It is therefore not implausible, at least in hindsight, that semi-classical gravity is able to calculate this entropy. 
\item A second and more perplexing feature is that the semi-classical calculation that leads to the unitarity-compatible Page curve involves the inclusion of replica wormholes into the Euclidean path integral \cite{Shenker, Malda}. When interpreted at face value, this suggests that we are in fact dealing with an {\em ensemble average}, when we use semi-classical gravity to compute the matrix elements that go into the entropy calculation \cite{Shenker}. Indeed for JT gravity in two dimensions, which is an ensemble average over unitary theories (and therefore is a non-unitary theory), one can explicitly demonstrate the emergence of the Page curve by evaluating the average over the underlying ensemble \cite{Saad, Shenker}. In short, {\em we seem to be finding a unitarity-compatible Page curve from an ensemble-averaged description}.
\end{itemize}

The second bullet point above, raises a puzzle. Our entire premise when looking for a tent-shaped (ie., unitarity-compatible) Page curve was that quantum gravity is unitary. And yet, now we have been dealt a devil's bargain. We have a unitarity-compatible Page curve, but in the semi-classical (Euclidean) gravity limit where we are working, it seems to be arising in an ensemble average over theories. Even though this is not quite a contradiction -- the ensemble average of a quantity that follows the Page curve will also follow the Page curve -- it does raise a puzzle about how one should think about the relationship between the fundamental description of gravity and its semi-classical description.

The Euclidean path integral is believed to be ill-defined as a complete definition of quantum gravity in higher dimensions (eg., the wrong sign kinetic term of the conformal mode of the metric). At the conceptual level, an obvious piece that is missing in our present understanding is the connection between semi-classical (bulk-metric based) gravity and the underlying ``true" quantum gravity degrees of freedom, which are presumably holographic. To make matters more confusing,  in low dimensions there seem to be non-unitary metric theories like JT gravity  that do have well-defined path integrals. These can be explicitly demonstrated to be ensemble averages over {\em distinct} unitary matrix models \cite{Saad}. 

The goal of this paper is to make some progress in understanding {\em how to think of semi-classical gravity} in more general contexts. More generally, we wish to understand the role (if any) of ensembles in a Page curve calculation in a unitary theory. Does the fact that semi-classical gravity is an ensemble average, suggest that the fundamental theory should also necessarily be an ensemble average over distinct theories? This is the case in JT gravity, and it has been suggested that this may be the general paradigm. Such an explicit ensemble average however would be disappointing from the point of view of the usual lore of the AdS/CFT correspondence, where {\em individual} unitary boundary theories (eg., ${\cal N}=4$ SYM) seem to be dual to {\em individual} unitary theories of quantum gravity (eg., type IIB string theory on AdS$_5 \times S^5$) which should each have semi-classical supergravity limits. We do expect black holes to arise as thermalized states in a {\em single copy} of an ${\mathcal N}=4$ SYM theory. 

% and this seems to be the philosophy suggested in \cite{Bousso}\footnote{See in particular, the discussion at the end of p. 6 in \cite{Bousso}.}

%Furthermore, it is not clear to us how to interpret the idea that we are living in an ensemble -- while each individual matrix model in the dual of JT gravity is a well-defined unitary theory, the 

%It has been suggested that this means that gravity should be viewed as an ensemble of unitary theories (eg., \cite{Bousso}). As a fundamental description of reality, an ensemble of theories clearly requires a paradigm shift. So our agenda in this paper will be to see whether some of the relevant questions can be understood in a more conservative setting, {\em not} complicated by horizons and the information paradox. 

%At first, this may raise the worry that we are throwing the baby out with the bathwater. After all, the origins of the recent discussions on the Page curve are to be found in attempts to resolve the black hole information paradox. 

In order to shed some light on this question, we will make two key observations in this paper regarding the two bullet points mentioned at the beginning of this Introduction. These observations are --
\begin{itemize}
\item Neither of the points have a {\em a priori} anything to do with gravity, black holes or horizons. By this we mean that both features can be seen in systems that apparently\footnote{It is an interesting question whether the systems we consider (eg., the hard sphere gas) are secretly dual to some (perhaps exotic) theory of gravity.} are without gravity.
\item Both features can be seen already at the level of {\em individual} unitary theories, without explicit ensemble averages. The ensembles arise much like they do in conventional statistical mechanics, where they arise as proxies for time averages when the system is in (approximate) thermal equilibrium.
\end{itemize}
In other words, the first bullet point about the semi-classical accessibility of the unitarity-compatible Page curve is equally valid in non-gravitational unitary theories. Similarly, there does not seem to be anything forbidding us  from coming up with a non-gravitational theory where a Page curve emerges at the semi-classical level via an apparent ensemble-average. Indeed, the bulk of this paper deals with the detailed study of an example that illustrates both these points. We expect that such examples should be fairly generically constructible in quantum chaotic systems which can be split into two subsystems.

% The claim then will be that this provides an existence argument for our two central taleaways: (1) the emergence of a unitarity-compatible Page curve in an ensemble averaged description is not a unique feature of gravity, and (2) even when the semi-classical description has features of an ensemble average, the full theory can be unitary. 

Our goal in the rest of the paper will be to exhibit these two ingredients in a single realization of a non-gravitational unitary theory. In our view, this strengthens the possibility that gravity may also fit into the same rubric: the semi-classical replica wormhole calculation reproduces the unitarity-compatible Page curve via an {\em apparent} ensemble average, while the full quantum gravity indeed remains safely unitary. Closely related ideas have appeared earlier, see \cite{Sully, Liu, Nomura1, Nomura2}. One of the new features in our calculation will be that we are able to follow the evolution of the system (and the Page curve) explicitly at the semi-classical level. This also enables us to have a clear understanding of the epoch-dependence of the ensemble. Other crucial features of our explicit model will become clear as we proceed.

If this picture is correct, low-dimensional examples like JT gravity which come with explicit ensemble averages and well-defined (but non-unitary) metric path integrals, are to be viewed as exceptions. The swampland ideas of \cite{Vafa}, which suggest that in high enough dimensions, ensembles for gravity contain only a {\em single} theory seem consistent with this picture. What is nice about JT gravity then, is that it gives us an explicitly doable, well-defined {\em metric} path integral unlike more realistic theories of gravity.
 
In what follows, we will work with the concrete example of a hard sphere gas leaking {\em slowly} from a small box into a larger one. A hard sphere gas in a box is known to be a quantum chaotic system, whose eigenstates were conjectured by Berry \cite{Berry1, Berry2} to behave semi-classically as though they were picked from a Gaussian ensemble. We will call this conjectural ensemble, Berry's ensemble. Berry's conjecture was one of the initial motivations for the Eigenstate Thermalization Hypothesis (ETH) \cite{Srednicki1}, see also \cite{Deutsch0}. The reason for our interest in this particular set up involving the hard sphere gas is that based on general principles of unitarity, we expect to see a Page curve in this system if we compute the entanglement entropy of the larger box. Equally importantly, thanks to Berry's conjecture, we may suspect (and indeed we will demonstrate) that it should be possible to show the emergence of this Page curve via a calculation at the  {\em semi-classical} level, where an {\em ensemble average} plays a significant role. 

%Our calculations and results turn out to be instructive for the black hole Page curve problem, despite the absence of horizons or a genuine information paradox. The Renyi entropies we calculate have some structural similarities to the replica wormholes calculation \cite{Shenker}. This is a consequence of the fact that 

Horizons, islands and other geometric objects do not play a  role in our calculations, and there is no genuine information paradox. But note that the questions we are interested in have only to do with the semi-classical ensemble average aspect, and we will show that our system shares that with the black hole system. Therefore, despite the differences, the lessons we extract from the hard sphere gas have a chance of holding for gravity as well. Indeed, this is our primary motivation behind the present paper. 

 % In particular, we do not have anything to say directly about islands or other intrinsically geometric objects -- our concern is only with the semi-classical ensemble average.

We will find that the semi-classical entanglement entropy of the larger box, follows the Page curve. The assumption of {\em slow} leakage\footnote{We will make this more precise later.}, leads to two timescales in the problem and we find that there is an analogue of a Hawking radiation {\em epoch} \cite{Penington} in the present problem as well. During each epoch, we can compute the entanglement entropy assuming that the eigenstates of the relevant subsystem are taken from Berry's ensemble\footnote{In our system, an epoch is characterized by the number of hard spheres in the larger box. This is conceptually parallel to how the clock used in \cite{Shenker} was the string of Hawking photons emitted up to that point.}. The result, when plotted against epoch, yields a unitarity-compatible Page curve. 

Interestingly enough, we also find that despite the absence of horizons, we have a simple way to obtain an information ``paradox" in this system. Instead of computing the ensemble-average of the Renyi entropy from the reduced density matrix, one can consider the Renyi entropy of the {\em ensemble-averaged} reduced density matrix. By direct calculation through the epochs, we find that the evolution of this object does not have the  turnaround and we are left with Page's version of the Hawking paradox. 

A key technical assumption in our calculation is that the leakage is slow so that the gas in each box can come to approximate equilibrium during each epoch. This is what enables us to take advantage of Berry's ensemble averaging epoch by epoch. In doing so, we are effectively assuming that the entanglement entropy during each epoch can be computed via a suitable time average (thanks to local equilibrium) and that the ensemble average is an ergodic stand-in for this, as is often the case in statistical mechanics. The entanglement entropy of the reduced density matrix of the larger system is the thermodynamic entropy of the {\em smaller} system during that epoch. In the limit when the system has fully thermalized and both boxes have the same density of particles, this reduces to the result obtained in \cite{Deutsch1}. So our work can be viewed as a type of generalization of the result there. See \cite{Deutsch} for some related discussions. 

The structure of these observations strongly suggest that perhaps a similar mechanism is what holds in gravity as well. %The fact that an evaporating black hole has a well-defined Hawking temperature, is equivalent to the assumption of local equilibrium in the dual holographic description. The timescale during which this temperature is approximately  constant is the relevant epoch in the black hole evaporation problem (its use was emphasized in eg., \cite{Penington}). 
By analogy with the hard sphere gas, we are therefore tempted to conjecture that semi-classical gravity is providing an ergodic ensemble averaged description of quantum gravitational dynamics in bulk local equilibrium.   Since gravity is holographic, more ideas will be needed to make this into a fully concrete proposal, but let us make one speculative comment. We suspect that some approximate notion of coarse-graining will likely be required in defining the relevant entanglement entropy in flat space gravity. A cut-off has played a role in flat space ever since the work of Gibbons-Hawking \cite{GH}, and it seems plausible to us that its correct interpretation is in implementing a coarse-graining \cite{More}. 

Our work departs from some of the statements in the literature, which call for gravity to be viewed as an explicit ensemble average. On the contrary, we view our results as being in line with the ideas of \cite{Sully, Liu}. Our purpose here is to present a concrete non-gravitational model which illustrates the relevant points, with an essentially fully calculable semi-classical Page curve. We believe this provides a clean context to evaluate the various ingredients, as well as the precise role played by gravity. We hope that our result is of some use in shedding light on how to think of the semi-classical gravity path integral, and figuring out its ultimate significance in the unitarity of the microscopic/holographic description of quantum gravity. 

%{\bf Until the densities match, the gas will keep leaking.} 

\section{Two Boxes for the Hard Sphere Gas}

Let us start by considering a collection of $N$ hard spheres, each with a radius $a$, enclosed in a cubic box of length\footnote{\label{4}In \cite{Srednicki1} the box size was taken to be $L+2a$. This makes sure that the centers of the spheres are living in a box of length $L$. This adds nothing to our discussion, and makes the definition of the hole connecting the two boxes slightly unwieldy, so we will let the centers themselves bounce off the box walls. This is purely a mathematical convenience.} $L$. Assume that there is a larger empty box of length $L^{\prime}$ in contact with the smaller box. At $t=0$, we open a hole in the wall between them so that the gas can leak slowly into the larger box. By tuning the size of the hole, we can take the leakage rate to be slow. We will take the size of the hole $d$ to be  somewhat larger than the sphere radius $a$ (there is some freedom in how big the hole can be chosen), so that the hole slows down the escape of particles from one box to the other. We will also take the average energy of the particles to be high enough so that approximate equilibrium can occur in each box during an $epoch$ (the timescale during which the macroscopic quantities associated to either box does not change appreciably). We would also like to work in the low density regime where simple calculations are possible \cite{Srednicki1}. With the mean free path $\ell = \tilde L^3/\sqrt{2}\pi \tilde N a^2$ (see eg., \cite{mfp}), it suffices to have
\bea
a \lesssim d \ll \frac{\tilde L^3}{\tilde Na^2} \label{eqslowleak}
\eea
Here $\tilde L$ denotes the fact that we are referring to either of the boxes, and $\tilde N$ is the number of particles in it during an epoch. 

We will model the system by assuming the hard spheres to be point particles/centers satisfying the constraint that the distance between any two centers cannot be less than $2a$. If all the particles were enclosed in a single box, this description will reduce to the model discussed in \cite{Srednicki1}. It is natural to expect our system with the two connected boxes also to exhibit ergodicity and chaos, even though typically the hard sphere gas in a single box is the one that is studied in the context of chaos and thermalization \cite{Srednicki1}. 

Let us look at the Hilbert space of the system. We can denote the energy eigenstates by $|\Psi_{\alpha}\ket$. Let us introduce a position basis $|\mathbf{X}\ket$, where $\mathbf{X}$ corresponds to the $3N$ dimensional position vector of all the particles. In this position basis, we can define the wavefunctions
\bea
\Psi{}_\alpha(\mathbf{X}) = \bra \mathbf{X}|\Psi_{\alpha}\ket \ \ \text{and} \ \   \Psi^{*}_{\alpha^{\prime}}(\mathbf{X}^{\prime}) =\bra \Psi_{ {\alpha^{\prime}}} |\mathbf{X}^{\prime}\ket. 
\eea
To define the domain where the wave function is defined, we first introduce an auxiliary domain 
\bea
D'=\left\{\mathbf{X}_{1}, \ldots, \mathbf{X}_{N}\ \Bigl{|} \ \mathbf{X}_{i} \in B_S \cup B_L ; \ |\mathbf{X}_{i}-\mathbf{X}_{j} \mid \geq 2 a\right\}
\eea
where the three Cartesian coordinates of the individual box domains are $B_S \equiv \left[0,L \right]^3$ and $B_L\equiv\left[L,L+L^{\prime} \right]\times\left[0,L^{\prime} \right]^2$.
The crucial extra boundary condition that defines the true domain of the system is given by the condition that the wavefunction vanishes not on all of $\partial B_1 \cup \partial B_2 $, but only on $\partial B_1 \cup \partial B_2  -H$ where $H$ is the part of the domain which corresponds to the location of the hole. The region within this vanishing condition of the wave function is our true domain, and we denote it by $D$. We will not need to specify the shape and location of the hole in detail to do our calculations below, other than the conditions on its size we noted above. Note that the second box is bigger than the first, ie.  $L' > L$, and $H$ is a subset of $\partial B_1 \cap \partial B_2$. As we will see later, in order to prevent the system from coming to global equilibrium before reaching the Page time, we can take $L' \gg L$. There is quite a bit of leeway in many of the choices we are making here.

The hierarchy in \eqref{eqslowleak} introduces two time-scales into the problem. Since the gas is leaking slowly, the time taken for each of the boxes to reach approximate equilibrium (separately), will be much smaller than the timescale of leakage during which the number of particles in the boxes change appreciably. Implicit is also the assumption that the average energies are sufficiently high that each box thermalizes quickly enough compared to the other scales in the problem. In any event, the end result is an epoch where both of the boxes have separately equilibrated and the number of particles in each of the boxes remains approximately fixed. Let $N_S$ and $N_L$ denote the number of particles in the smaller and larger box at a particular epoch. As the total number of hard spheres in the boxes remain fixed throughout an epoch, we can use either $N_S$ or $N_L$ to characterize it. The number of hard spheres are related to each other through the conservation law
\bea
N_S+N_L =N. \label{eqparticlenumber}
\eea

To study the evolution of the state of the system, we expand it as a linear combination of the eigenstates $|\Psi_{\alpha}\ket$ of the full system (ie., the two boxes connected by the hole) as follows:
\bea
|\Psi\ket  = \sum_{\alpha} d^{}_{\alpha}|\Psi_{\alpha}\ket
\eea
The corresponding density matrix of the state is
\bea
\rho = |\Psi\ket\bra \Psi| = \sum_{\alpha,{\alpha^{\prime}}} d^{}_{\alpha}d^{*}_{ {\alpha^{\prime}}}|\Psi_{\alpha}\ket\bra\Psi_{ {\alpha^{\prime}}}|
\eea
In the position basis, we have
\bea
\rho(\mathbf{X};\mathbf{X}^{\prime}) \equiv \bra \mathbf{X}|\rho |\mathbf{X}^{\prime}\ket = \Psi(\mathbf{X})  \Psi^{*}(\mathbf{X}^{\prime})=\sum_{\alpha,{\alpha^{\prime}}} d^{}_{\alpha}d^{*}_{ {\alpha^{\prime}}}\ \Psi{}_\alpha(\mathbf{X})  \Psi^{*}_{\alpha^{\prime}}(\mathbf{X}^{\prime}), \label{eqfullstate1}
\eea
where $\Psi(\mathbf{X})\equiv \bra \mathbf{X}|\Psi\ket$. In order to analyze the properties of each of the boxes separately with the epochs, it becomes useful to split up the coordinates in a convenient epoch-dependent manner. At every epoch, we have \eqref{eqparticlenumber} and this provides a natural partition of the $3N$ components of the vector $\mathbf{X}$ into $3N_S$ and $3N_L$ components as follows:
\bea
\mathbf{X} = (\mathbf{x},\mathbf{y}) \ \ \text{and} \ \ |\mathbf{X}\ket = |\mathbf{x},\mathbf{y}\ket
\eea
where $\mathbf{x} = (\mathbf{x}_{1}, \ldots, \mathbf{x}_{N_S})$ and $\mathbf{y} = (\mathbf{y}_{1}, \ldots, \mathbf{y}_{N_L})$ can  loosely be thought of as denoting the position vectors of the particles in the smaller and larger boxes respectively\footnote{But note however that at this stage, the ranges of the positions of each of the particles span the full system, and not the left or right box alone. A general quantum state in the Hilbert space is an arbitrary $superposition$ of states with no particular notion of localization. Note that viewing the particles as being localized in one of the boxes is a kind of semi-classical approximation. We will make such an assumption eventually, and the notation is introduced with that in mind.}. 
%\footnote{This partition has the same structure as the decomposition of a three-dimensional position vector into its Cartesian components $\mathbf{x} = (x_1,x_2,x_3)$ and $|\mathbf{x}\ket = |x_1,x_2,x_3\ket$}. 
In terms of these coordinates, we can define the wavefunctions as
\bea
\Psi{}_\alpha(\mathbf{x},\mathbf{y}) = \bra\mathbf{x},\mathbf{y}|\Psi_{\alpha}\ket \ \ \text{and} \ \   \Psi^{*}_{\alpha^{\prime}}(\mathbf{x},\mathbf{y}) =\bra \Psi_{ {\alpha^{\prime}}} |\mathbf{x},\mathbf{y}\ket 
\eea
In this notation for the position basis, the density matrix takes the form 
\bea
\rho(\mathbf{x},\mathbf{y};\mathbf{x}^{\prime},\mathbf{y}^{\prime}) \equiv \bra\mathbf{x},\mathbf{y}|\rho |\mathbf{x}^{\prime},\mathbf{y}^{\prime}\ket = \sum_{\alpha,{\alpha^{\prime}}} d^{}_{\alpha}d^{*}_{ {\alpha^{\prime}}}\ \Psi{}_\alpha(\mathbf{x},\mathbf{y})  \Psi^{*}_{\alpha^{\prime}}(\mathbf{x}^{\prime},\mathbf{y}^{\prime}) 
\eea
\section{Purity of the Larger Box}
\label{puritysec}
\noindent To calculate the entanglement entropy of the larger box, we will compute the $n$-th Renyi entropy of subsystem, and then continue to $n \to 1$. As a warm-up, we will start with the computation of the purity ($n=2$) of the larger box. We start by computing the reduced density matrix of the particles ``associated to the larger box", in the notation of the last paragraphs of the previous section:
\bea
\begin{aligned}
\rho_{L}(\mathbf{y};\mathbf{y}^{\prime}) =\int_{D} dx  \ \rho(\mathbf{x},\mathbf{y} ;  \mathbf{x},\mathbf{y}^{\prime})  
= \int_{D} dx \sum_{\alpha,{\alpha^{\prime}}} d^{}_{\alpha}d^{*}_{ {\alpha^{\prime}}} \ \Psi_\alpha(\mathbf{x},\mathbf{y})  \Psi^{*}_{\alpha^{\prime}}(\mathbf{x},\mathbf{y}^{\prime})
\end{aligned}\label{eqreduceddensity}
\eea
Squaring the matrix, we get
\bea
\rho^{2}_{L}(\mathbf{y};\mathbf{y}^{\prime}) &=& \int_{D} dy^{\prime \prime}  \ \rho_{L}(\mathbf{y};\mathbf{y}^{\prime \prime})\rho_{L}(\mathbf{y}^{\prime \prime};\mathbf{y}^{\prime})  \\
&=& \int_{D} dy^{\prime \prime} dx dx^{\prime}\sum_{\alpha^{}_{1},{\alpha^{\prime}_{1}},\alpha^{}_{2},{\alpha^{\prime}_{2}}} d^{}_{\alpha^{}_{1}}d^{*}_{ {\alpha^{\prime}_{1}}}d^{}_{{\alpha^{}_{2}}}d^{*}_{ {\alpha^{\prime}_{2}}}  \ \Psi_{\alpha^{}_{1}}(\mathbf{x},\mathbf{y})  \Psi^{*}_{\alpha^{\prime}_{1}}(\mathbf{x},\mathbf{y}^{\prime \prime})\Psi_{\alpha^{}_{2}}(\mathbf{x}^{\prime},\mathbf{y}^{\prime \prime})  \Psi^{*}_{\alpha^{\prime}_{2}}(\mathbf{x}^{\prime},\mathbf{y}^{\prime}) \hspace{0.2cm}\nonumber \label{eqrho2}
\eea
Now let us look at the behavior of this quantity at each epoch. From the discussion in the previous section, we can see that working in various epochs is equivalent to restricting ourselves to processes occurring at time-scales larger than the equilibrization time of each box. Therefore, we can effectively replace the relevant quantities with their time averages over this timescale. As we expect the system to be ergodic, we can in turn replace the time average with an ensemble average. Therefore, to understand the behavior of quantities in each epoch, we should look at their averages in the appropriate ensemble\footnote{We will assume that the Renyi and entanglement entropies are quantities that can be calculated in this way. We will also assume that the (eigenstate) ensemble replacement can be done when the system is in $local$ equilibirum. The fact that the results are reasonable (as we will see) will be taken as $a$ $posteriori$ evidence for these assumptions.}.

It turns out that there is a natural choice for such an ensemble. Consider a quantum chaotic system. Berry's conjecture  \cite{Berry1, Berry2} says that when the energy of an eigenstate is sufficiently high, the state behaves as if it was picked randomly from a fictitious Gaussian ensemble. It was shown in \cite{Srednicki1} that when evaluated in this eigenstate ensemble\footnote{We will refer to this ensemble as Berry's ensemble in the context of the hard sphere gas.}, the single particle momentum distribution function of the hard sphere gas turned out to be equal to the Maxwell-Boltzmann distribution. This is a specific manifestation of the \textit{eigenstate thermalization hypothesis} (ETH). It is expected that (see eg., \cite{Deutsch}) for systems which satisfy the ETH condition, ergodicity is guaranteed. Therefore, we can hope that averaging over Berry's ensemble acts as an ergodic proxy for the underlying time averaging. A further comment worth making, is that Berry's conjecture is based on semi-classical physics and relies on the connection between classical and quantum chaos \cite{review}. So this further strengthens the parallel with the black hole Page curve calculation, which was done in the setting of semi-classical gravity \cite{Shenker}. 

Adopting this philosophy, we are now ready to compute the purity of the reduced density matrix in Berry's ensemble:
%\bea \begin{aligned} \left\langle\rho^{2}_{L}(\mathbf{y};\mathbf{y}^{\prime}) \right\rangle_{\mathrm{EE}} = \int_{D{}_L} dy^{\prime \prime} \int_{D{}_S} dx dx^{\prime}\sum_{\alpha^{}_{1},{\alpha^{\prime}_{1}},\alpha^{}_{2},{\alpha^{\prime}_{2}}}  &d^{}_{\alpha^{}_{1}}d^{*}_{ {\alpha^{\prime}_{1}}}d^{}_{{\alpha^{}_{2}}}d^{*}_{ {\alpha^{\prime}_{2}}} \\ & \left\langle\Psi_{\alpha^{}_{1}}(\mathbf{x},\mathbf{y})  \Psi^{*}_{\alpha^{\prime}}(\mathbf{x},\mathbf{y}^{\prime \prime})\Psi_{\alpha^{}_{2}}(\mathbf{x}^{\prime},\mathbf{y}^{\prime \prime})  \Psi^{*}_{\alpha^{\prime}_{2}}(\mathbf{x}^{\prime},\mathbf{y}^{\prime})\right\rangle_{\mathrm{EE}}\end{aligned} \eea
% Therefore, the purity of the larger box in the eigenstate ensemble will be given by
\bea
\begin{aligned}
&\text{Tr}\left\langle\rho^{2}_{L} \right\rangle_{\mathrm{EE}}\\
& = \int_{D} dy \left\langle\rho^{2}_{L}(\mathbf{y};\mathbf{y}) \right\rangle_{\mathrm{EE}}\\
&=  \int_{D} dy^{\prime \prime}  dy \int_{D} dx dx^{\prime}\sum_{\alpha^{}_{1},{\alpha^{\prime}_{1}},\alpha^{}_{2},{\alpha^{\prime}_{2}}}  d^{}_{\alpha^{}_{1}}d^{*}_{ {\alpha^{\prime}_{1}}}d^{}_{{\alpha^{}_{2}}}d^{*}_{ {\alpha^{\prime}_{2}}} \left\langle\Psi_{\alpha^{}_{1}}(\mathbf{x},\mathbf{y})  \Psi^{*}_{\alpha^{\prime}_1}(\mathbf{x},\mathbf{y}^{\prime \prime})\Psi_{\alpha^{}_{2}}(\mathbf{x}^{\prime},\mathbf{y}^{\prime \prime})  \Psi^{*}_{\alpha^{\prime}_{2}}(\mathbf{x}^{\prime},\mathbf{y})\right\rangle_{\mathrm{EE}}\label{pureq} 
\end{aligned}
\eea
where the subscript $\mathrm{EE}$ denotes that the quantity is averaged over the eigenstate ensemble. Berry's conjecture  would imply that the four-point function\footnote{Note that what we are evaluating is actually a product of eigenfunctions, not a correlation function. But we will use this slightly distracting terminology because of the obvious parallel in structure, and to $not$ keep repeating the lengthy phrase ``ensemble expectation value of the product of eigenfunctions".} will be given in terms of the Wick contractions of the two-point functions, as in \cite{Srednicki1} (see also appendix \ref{npointapp}, for related discussions in the single box). Therefore, we have
\bea
\begin{aligned}
\Bigl{\langle}\Psi_{\alpha^{}_{1}}(\mathbf{x},\mathbf{y})  \Psi^{*}_{\alpha^{\prime}_1}(\mathbf{x},\mathbf{y}^{\prime \prime})&\Psi_{\alpha^{}_{2}}(\mathbf{x}^{\prime},\mathbf{y}^{\prime \prime})  \Psi^{*}_{\alpha^{\prime}_{2}}(\mathbf{x}^{\prime},\mathbf{y})\Bigl{\rangle}_{\mathrm{EE}}\\
& =\Bigl{[} \left\langle\Psi^{}_{\alpha^{}_{1}}(\mathbf{x},\mathbf{y})  \Psi^{*}_{\alpha^{\prime}_{1}}(\mathbf{x},\mathbf{y}^{\prime \prime})\right\rangle_{\mathrm{EE}} \left\langle\Psi_{\alpha^{}_{2}}(\mathbf{x}^{\prime},\mathbf{y}^{\prime \prime})  \Psi^{*}_{\alpha^{\prime}_{2}}(\mathbf{x}^{\prime},\mathbf{y})\right\rangle_{\mathrm{EE}} \\
&+ \left\langle\Psi^{}_{\alpha^{}_{1}}(\mathbf{x},\mathbf{y})  \Psi_{\alpha^{}_{2}}(\mathbf{x}^{\prime},\mathbf{y}^{\prime \prime})\right\rangle_{\mathrm{EE}} \left\langle \Psi^{*}_{\alpha^{\prime}_{1}}(\mathbf{x},\mathbf{y}^{\prime \prime}) \Psi^{*}_{\alpha^{\prime}_{2}}(\mathbf{x}^{\prime},\mathbf{y})\right\rangle_{\mathrm{EE}} \\
&+ \left\langle\Psi^{}_{\alpha^{}_{1}}(\mathbf{x},\mathbf{y}) \Psi^{*}_{\alpha^{\prime}_{2}}(\mathbf{x}^{\prime},\mathbf{y})\right\rangle_{\mathrm{EE}} \left\langle \Psi^{*}_{\alpha^{\prime}_{1}}(\mathbf{x},\mathbf{y}^{\prime \prime})   \Psi_{\alpha^{}_{2}}(\mathbf{x}^{\prime},\mathbf{y}^{\prime \prime})\right\rangle_{\mathrm{EE}} \Bigl{]} \label{purity8pointeq}
\end{aligned}
\eea
Plugging the above expression into the previous one, we get
\bea
\begin{aligned}
\text{Tr}\left\langle\rho^{2}_{L} \right\rangle_{\mathrm{EE}}& = \int_{D} dy^{\prime \prime}  dy \int_{D} dx dx^{\prime} \sum_{\alpha^{}_{1},{\alpha^{\prime}_{1}},\alpha^{}_{2},{\alpha^{\prime}_{2}}}  d^{}_{\alpha^{}_{1}}d^{*}_{ {\alpha^{\prime}_1}}d^{}_{{\alpha^{}_{2}}}d^{*}_{ {\alpha^{\prime}_{2}}} \\
&\hspace{2cm} \Bigl{[} \left\langle\Psi^{}_{\alpha^{}_{1}}(\mathbf{x},\mathbf{y})  \Psi^{*}_{\alpha^{\prime}_{1}}(\mathbf{x},\mathbf{y}^{\prime \prime})\right\rangle_{\mathrm{EE}} \left\langle\Psi_{\alpha^{}_{2}}(\mathbf{x}^{\prime},\mathbf{y}^{\prime \prime})  \Psi^{*}_{\alpha^{\prime}_{2}}(\mathbf{x}^{\prime},\mathbf{y})\right\rangle_{\mathrm{EE}} \\
&\hspace{2cm} + \left\langle\Psi^{}_{\alpha^{}_{1}}(\mathbf{x},\mathbf{y})  \Psi_{\alpha^{}_{2}}(\mathbf{x}^{\prime},\mathbf{y}^{\prime \prime})\right\rangle_{\mathrm{EE}} \left\langle \Psi^{*}_{\alpha^{\prime}_{1}}(\mathbf{x},\mathbf{y}^{\prime \prime}) \Psi^{*}_{\alpha^{\prime}_{2}}(\mathbf{x}^{\prime},\mathbf{y})\right\rangle_{\mathrm{EE}} \\
&\hspace{2cm} + \left\langle\Psi^{}_{\alpha^{}_{1}}(\mathbf{x},\mathbf{y}) \Psi^{*}_{\alpha^{\prime}_{2}}(\mathbf{x}^{\prime},\mathbf{y})\right\rangle_{\mathrm{EE}} \left\langle \Psi^{*}_{\alpha^{\prime}_{1}}(\mathbf{x},\mathbf{y}^{\prime \prime})   \Psi_{\alpha^{}_{2}}(\mathbf{x}^{\prime},\mathbf{y}^{\prime \prime})\right\rangle_{\mathrm{EE}} \Bigl{]}\\
\end{aligned}
\eea
Pulling the sums into the ensemble average, this becomes
\bea
\begin{aligned}
\text{Tr}\left\langle\rho^{2}_{L} \right\rangle_{\mathrm{EE}}= \int_{D} dy^{\prime \prime}  dy \int_{D} dx dx^{\prime} \ &\Bigl{[} \left\langle\Psi(\mathbf{x},\mathbf{y})  \Psi^{*}(\mathbf{x},\mathbf{y}^{\prime \prime})\right\rangle_{\mathrm{EE}} \left\langle\Psi(\mathbf{x}^{\prime},\mathbf{y}^{\prime \prime})  \Psi^{*}(\mathbf{x}^{\prime},\mathbf{y})\right\rangle_{\mathrm{EE}} \\
& + \left\langle\Psi^{}(\mathbf{x},\mathbf{y})  \Psi(\mathbf{x}^{\prime},\mathbf{y}^{\prime \prime})\right\rangle_{\mathrm{EE}} \left\langle \Psi^{*}(\mathbf{x},\mathbf{y}^{\prime \prime}) \Psi^{*}(\mathbf{x}^{\prime},\mathbf{y})\right\rangle_{\mathrm{EE}} \\
& + \left\langle\Psi^{}(\mathbf{x},\mathbf{y}) \Psi^{*}(\mathbf{x}^{\prime},\mathbf{y})\right\rangle_{\mathrm{EE}} \left\langle \Psi^{*}(\mathbf{x},\mathbf{y}^{\prime \prime})   \Psi(\mathbf{x}^{\prime},\mathbf{y}^{\prime \prime})\right\rangle_{\mathrm{EE}} \Bigl{]}\\
\end{aligned}
\eea

The above expression (and its natural higher Renyi generalizations) will be the starting point for our calculations\footnote{We strongly suspect that the replacement of products of wave functions by suitable Gaussian ensemble expectations is a general property of quantum states that have a macroscopic interpretation as being in local equilibrium. This is a more general claim than Berry's conjecture. The latter, from this perspective, is a corollary of this claim together with the expectation that high lying eigenstates are (global) equlibrium states.}. Now let us evaluate the two-point functions in the above expression. At each epoch, we are making a semi-classical approximation that $N_S$ particles are in one box and the rest are in the other. At the level of wave functions, this enables us to assume that the value of the wavefunction $\Psi$ vanishes (at least approximately) at the hole $H$. Roughly, at each epoch, we choose the boundary condition that $\Psi$ vanishes on the boundary of $D_S$ and $D_L$ where these domains characterize the two separate boxes (and are defined precisely below). So we can decompose the state $\Psi$ as follows:
\bea
\Psi(\mathbf{x},\mathbf{y}) \approx  \sum_{i^{}_S,i^{}_L} c^{}_{i^{}_Si^{}_L}\psi_{i^{}_S}(\mathbf{x})\phi_{i^{}_L}(\mathbf{y}) \label{eqfactor}
\eea
where $\psi_{i^{}_S}(\mathbf{x})$ and $\phi_{i^{}_L}(\mathbf{y})$ are the eigenfunctions of the smaller and larger boxes, with $N_S$ and $N_L$ hard spheres respectively. These wavefunctions are defined in the domains $D_S$ and $D_L$
%The domain of these wavefunctions also admits a natural split: \bea D = D_L \cup D_S \eea
where
\bea
D_S=\left\{\mathbf{x}_{1}, \ldots, \mathbf{x}_{N_S}\ \Bigl{|} \ \mathbf{x}_i \in \left[0,L \right]^3; \ |\mathbf{x}_{i}-\mathbf{x}_{j} \mid \geq 2 a\right\}
\eea
and 
\bea
D_L=\left\{\mathbf{y}_{1}, \ldots, \mathbf{y}_{N_L} \ \Bigl{|} \ \mathbf{y}_{i} \in \left[L,L+L^{\prime} \right]\times\left[0,L^{\prime} \right]^2  ; \ |\mathbf{y}_{i}-\mathbf{y}_{j} \mid \geq 2 a\right\}
\eea
and they vanish on the boundary of their respective domains. 
 
Eqn (\ref{eqfactor}) is our semi-classical approximation for the state in a particular epoch defined by the number of particles in each box. It is motivated by the slow leakage assumption and  the expectation of approximate equilibrium in each box during an epoch. Because of the latter, we expect that the support for the state from the eigenstates in each box will come in a narrow sliver. Let us make this precise by making a few definitions. From the reduced density matrices, we can calculate the average energy of boxes:
\bea
\bar{U}_L =  \sum_{i^{}_{{S}},i^{}_{{L}}} \left| c^{}_{ i^{}_{{S}}i^{}_{{L}}}\right|^2 U_{i^{}_{L}} \ \ \text{and} \ \ \bar{U}_S =  \sum_{i^{}_{{S}},i^{}_{{L}}} \left| c^{}_{ i^{}_{{S}}i^{}_{{L}}}\right|^2 U_{i^{}_{S}}\label{varianceeq}
\eea 
where $U_{i^{}_{S}}$ and $U_{i^{}_{L}}$ correspond to the energy eigenvalues of the smaller and larger boxes respectively. The uncertainty in the average energy is given by
\bea
{\Delta}_{L}^{2}=\sum_{i^{}_{{S}},i^{}_{{L}}} \left| c^{}_{ i^{}_{{S}}i^{}_{{L}}}\right|^2 \left(U_{i^{}_{L}}-\bar{U}_{L}\right)^{2}  \ \ \text{and} \ \ {\Delta}_{S}^{2}=\sum_{i^{}_{{S}},i^{}_{{L}}} \left| c^{}_{ i^{}_{{S}}i^{}_{{L}}}\right|^2 \left(U_{i^{}_{S}}-\bar{U}_{S}\right)^{2}.
\eea 
We can also define the temperature of the boxes at each epoch using the average energies:
\bea
\bar{U}_L \equiv \frac{3}{2}N_L k \bar{T}_L \ \ \text{and} \ \ \bar{U}_S \equiv \frac{3}{2}N_S k \bar{T}_S
\eea
These definitions follow \cite{Srednicki1} and will be useful to us in our calculations.

%In other words  these are eigenfunctions for particles in the isolated small/large boxes, respectively. 
The expression for the purity with the semi-classical state can be written somewhat schematically in the form
\bea
\begin{aligned}
&\text{Tr}\left\langle\rho^{2}_{L} \right\rangle_{\mathrm{EE}} = \int_{D{}_L} dy^{\prime \prime}  dy \int_{D{}_S} dx dx^{\prime}\sum_{i^{}_{1_{S}},i^{\prime}_{1_S},i^{}_{1_{L}},i^{\prime}_{1_L},i^{}_{2_{S}},i^{\prime}_{2_S},i^{}_{2_{L}},i^{\prime}_{2_L}}  c^{}_{i^{}_{1_{S}}i^{}_{1_{L}}}c^{*}_{ i^{\prime}_{1_S}i^{\prime}_{1_L}} c^{}_{ i^{}_{2_{S}}i^{}_{2_{L}}}c^{*}_{ i^{\prime}_{2_S}i^{\prime}_{2_L}}\\
&\hspace{2cm} \Bigl{[}\left\langle\psi^{*}_{ i^{\prime}_1}(\mathbf{x})\phi^{*}_{ i^{\prime}_{1}}(\mathbf{y})\psi_{i^{}_{1} }(\mathbf{x})\phi_{ i^{}_{1}}(\mathbf{y}^{\prime \prime})\right\rangle_{\mathrm{EE}} \left\langle\psi^{*}_{i^{\prime}_{2}}(\mathbf{x^{\prime }})\phi^{*}_{ i^{\prime}_{2}}(\mathbf{y}^{\prime \prime})\psi_{i^{}_{2}}(\mathbf{x^{\prime }}) \phi_{ i^{}_{2}}(\mathbf{y})\right\rangle_{\mathrm{EE}} \\
&\hspace{2cm} +\left\langle\psi^{*}_{ i^{\prime}_1}(\mathbf{x})\phi^{*}_{ i^{\prime}_{1}}(\mathbf{y})\psi^{*}_{i^{\prime}_{2}}(\mathbf{x^{\prime }})\phi^{*}_{ i^{\prime}_{2}}(\mathbf{y}^{\prime \prime})\right\rangle_{\mathrm{EE}} \left\langle\psi_{i^{}_{1} }(\mathbf{x})\phi_{ i^{}_{1}}(\mathbf{y}^{\prime \prime})\psi_{i^{}_{2}}(\mathbf{x^{\prime }}) \phi_{ i^{}_{2}}(\mathbf{y})\right\rangle_{\mathrm{EE}} \\
&\hspace{2cm} +\left\langle\psi^{*}_{ i^{\prime}_1}(\mathbf{x})\phi^{*}_{ i^{\prime}_{1}}(\mathbf{y})\psi_{i^{}_{2}}(\mathbf{x^{\prime }}) \phi_{ i^{}_{2}}(\mathbf{y})\right\rangle_{\mathrm{EE}} \left\langle\psi^{*}_{i^{\prime}_{2}}(\mathbf{x^{\prime }})\phi^{*}_{ i^{\prime}_{2}}(\mathbf{y}^{\prime \prime})\psi_{i^{}_{1} }(\mathbf{x})\phi_{ i^{}_{1}}(\mathbf{y}^{\prime \prime})\right\rangle_{\mathrm{EE}}\Bigl{]}
\end{aligned}\label{traceeq1}
\eea
As $\mathbf{x}$ and $\mathbf{y}$ are independent variables, we can again simplify the expression in the square brackets using Berry's conjecture, now for the two-point functions in the individual boxes. This gives us
\bea
\begin{aligned}
&\text{Tr}\left\langle\rho^{2}_{L} \right\rangle_{\mathrm{EE}} = \int_{D{}_L} dy^{\prime \prime}  dy \int_{D{}_S} dx dx^{\prime}\sum_{i^{}_{1_{S}},i^{\prime}_{1_S},i^{}_{1_{L}},i^{\prime}_{1_L},i^{}_{2_{S}},i^{\prime}_{2_S},i^{}_{2_{L}},i^{\prime}_{2_L}}  c^{}_{i^{}_{1_{S}}i^{}_{1_{L}}}c^{*}_{ i^{\prime}_{1_S}i^{\prime}_{1_L}} c^{}_{ i^{}_{2_{S}}i^{}_{2_{L}}}c^{*}_{ i^{\prime}_{2_S}i^{\prime}_{2_L}}\\
&\hspace{2cm} \Bigl{[}\left\langle\psi^{*}_{ i^{\prime}_1}(\mathbf{x})\psi_{i^{}_{1} }(\mathbf{x})\right\rangle_{\mathrm{EE}}\left\langle\phi^{*}_{ i^{\prime}_{1}}(\mathbf{y})\phi_{ i^{}_{1}}(\mathbf{y}^{\prime \prime})\right\rangle_{\mathrm{EE}} \left\langle\psi^{*}_{i^{\prime}_{2}}(\mathbf{x^{\prime }})\psi_{i^{}_{2}}(\mathbf{x^{\prime }})\right\rangle_{\mathrm{EE}}\left\langle\phi^{*}_{ i^{\prime}_{2}}(\mathbf{y}^{\prime \prime}) \phi_{ i^{}_{2}}(\mathbf{y})\right\rangle_{\mathrm{EE}} \\
&\hspace{2cm} +\left\langle\psi^{*}_{ i^{\prime}_1}(\mathbf{x})\psi^{*}_{i^{\prime}_{2}}(\mathbf{x^{\prime }})\right\rangle_{\mathrm{EE}}\left\langle\phi^{*}_{ i^{\prime}_{1}}(\mathbf{y})\phi^{*}_{ i^{\prime}_{2}}(\mathbf{y}^{\prime \prime})\right\rangle_{\mathrm{EE}} \left\langle\psi_{i^{}_{1} }(\mathbf{x})\psi_{i^{}_{2}}(\mathbf{x^{\prime }})\right\rangle_{\mathrm{EE}}\left\langle\phi_{ i^{}_{1}}(\mathbf{y}^{\prime \prime}) \phi_{ i^{}_{2}}(\mathbf{y})\right\rangle_{\mathrm{EE}} \\
&\hspace{2cm}+\left\langle\psi^{*}_{ i^{\prime}_1}(\mathbf{x})\psi_{i^{}_{2}}(\mathbf{x^{\prime }}) \right\rangle_{\mathrm{EE}}\left\langle\phi^{*}_{ i^{\prime}_{1}}(\mathbf{y})\phi_{ i^{}_{2}}(\mathbf{y})\right\rangle_{\mathrm{EE}} \left\langle\psi^{*}_{i^{\prime}_{2}}(\mathbf{x^{\prime }})\psi_{i^{}_{1} }(\mathbf{x})\right\rangle_{\mathrm{EE}}\left\langle\phi^{*}_{ i^{\prime}_{2}}(\mathbf{y}^{\prime \prime})\phi_{ i^{}_{1}}(\mathbf{y}^{\prime \prime})\right\rangle_{\mathrm{EE}}\Bigl{]}
\end{aligned}\label{traceeq3}
\eea
The two-point functions are known/calculable \cite{Srednicki1} and are discussed in Appendix A. Using \eqref{eqZcondition}, we can do the above integrals, and we find 
\bea
\begin{aligned}
&\text{Tr} \left\langle \rho^{2}_{L} \right\rangle_{\mathrm{EE}} \\
&=\sum_{i^{}_{1_{S}},i^{\prime}_{1_S},i^{}_{1_{L}},i^{\prime}_{1_L},i^{}_{2_{S}},i^{\prime}_{2_S},i^{}_{2_{L}},i^{\prime}_{2_L}}  c^{}_{i^{}_{1_{S}}i^{}_{1_{L}}}c^{*}_{ i^{\prime}_{1_S}i^{\prime}_{1_L}} c^{}_{ i^{}_{2_{S}}i^{}_{2_{L}}}c^{*}_{ i^{\prime}_{2_S}i^{\prime}_{2_L}} \\
&\hspace{2cm}\Bigl{[}  \delta_{ i^{\prime}_{1_S},i^{}_{1_S} } \ \delta_{ i^{\prime}_{1_L},i^{}_{1_L} } \ \delta_{i^{\prime}_{2_S},i^{}_{2_S} } \ \delta_{ i^{\prime}_{2_L},i^{}_{2_L} }Z_{i^{}_{1_L}}\mathcal{S}\left( U_{i^{}_{1_L} },U_{i^{\prime}_{2_L}},L^{\prime}\right)\\
&\hspace{2cm}+\delta_{ i^{\prime}_{1_S},i^{\prime}_{2_S} } \ \delta_{ i^{\prime}_{1_L}, i^{\prime}_{2_L} } \ \delta_{i^{}_{1_S},i^{}_{2_S} } \ \delta_{i^{}_{1_L},i^{}_{2_L} } \ Z_{i^{}_{1_S}}Z_{ i^{}_{1_L}}\mathcal{S}\left( U_{i^{}_{1_L} },U_{i^{\prime}_{2_L}},L^{\prime}\right)\mathcal{S}\left(U_{i^{}_{S}},U_{i^{\prime}_{1_S}},L\right)\\
&\hspace{2cm}+\delta_{ i^{\prime}_{1_S},i^{}_{2_S} } \ \delta_{ i^{\prime}_{1_L},i^{}_{2_L} } \ \delta_{i^{\prime}_{2_S}, i^{}_{1_S}} \ \delta_{ i^{\prime}_{2_L},i^{}_{1_L} } \  Z_{i^{}_{1_S}}\mathcal{S}\left(U_{i^{}_{S}},U_{i^{\prime}_{1_S}},L\right)\Bigl{]} 
\end{aligned}
\eea
where we have defined
\bea
Z_{i^{}_{L}} = \left(\frac{L^{\prime}}{h}\right)^{-3N_{L}}\frac{\Gamma(3N_{L}/ 2) (2 m U_{i^{}_{L}})}{(2 \pi m U_{i^{}_{L}})^{3N_{L} / 2}} \ \  \ \ \  \ \  Z_{i^{}_{S}} = \left(\frac{L}{h}\right)^{-3N_{S}}\frac{\Gamma(3N_{S}/ 2) (2 m U_{i^{}_{S}})}{(2 \pi m U_{i^{}_{S}})^{3N_{S} / 2}}
\eea
and 
\bea
\mathcal{S}\left(U_{i },U_{j},\mathtt{L}\right) = \exp \left[\frac{-m\left(U_{i }-U_{j}\right)^{2}\mathtt{L}^{2}}{  8 \pi \hbar^{2} U_{i}}\right].
\eea
Simplifying the expression, we get
\bea
\begin{aligned}
&\text{Tr} \left\langle \rho^{2}_{L} \right\rangle_{\mathrm{EE}} \\
&=\sum_{i^{}_{1_{S}},i^{}_{1_{L}},i^{}_{2_{S}},i^{}_{2_{L}}} \left| c^{}_{ i^{}_{1_{S}}i^{}_{1_{L}}}\right|^2 \left| c^{}_{ i^{}_{2_{S}}i^{}_{2_{L}}}\right|^2 Z_{i^{}_{1_L}} \ \mathcal{S}\left( U_{i^{}_{1_L} },U_{i^{}_{2_L}},L^{\prime}\right) \\
&+\sum_{i^{}_{1_{S}},i^{}_{1_{L}},i^{\prime}_{1_{S}},i^{\prime}_{1_{L}}} \left| c^{}_{ i^{}_{1_{S}}i^{}_{1_{L}}}\right|^2 \left| c^{}_{ i^{\prime}_{1_{S}}i^{\prime}_{1_{L}}}\right|^2 Z_{i^{}_{1_S}}Z_{ i^{}_{1_L}} \ \mathcal{S}\left( U_{i^{}_{1_L} },U_{i^{\prime}_{1_L}},L^{\prime}\right)  \mathcal{S}\left(U_{i^{}_{S}},U_{i^{\prime}_{1_S}},L\right)\\
&+\sum_{i^{}_{1_{S}},i^{}_{1_{L}},i^{\prime}_{1_{S}},i^{\prime}_{1_{L}}} \left|c^{}_{ i^{}_{1_{S}}i^{}_{1_{L}}}\right|^2\left|c^{}_{ i^{\prime}_{1_{S}}i^{\prime}_{1_{L}}}\right|^2  \  Z_{i^{}_{1_S}} \ \mathcal{S}\left(U_{i^{}_{S}},U_{i^{\prime}_{1_S}},L\right)  \label{traceeq000}
\end{aligned}
\eea
Let us look at the first sum on the RHS of the above equation as a representative for how to massage this expression. From the the definition of $\mathcal{S}$, we can see that the sum will be dominated by terms with $\left|U_{i^{}_{1_{L}}}-U_{i^{}_{2_{L}}}\right| / U_{i^{}_{1_{L}}} \lesssim \left(\hbar^{2} / m U_{\alpha} {L^{\prime}}^{2}\right)^{1 / 2}$. To simplify the notation, let us define
\bea
\Delta^{\prime}_{L} = \bar{U}_L  \bar{\lambda}_L / N_L^{1 / 2} L^{\prime} \label{eqemergedelta1}
\eea
where $\bar{\lambda}_L=\left(2 \pi \hbar^{2} / m k \bar{T}_L\right)^{1 / 2}$ denotes the thermal wavelength at the temperature $\bar{T}_L$. It turns out that the discussion is simpler if we assume $\Delta_L \lesssim \Delta^{\prime}_{L}$. In that case, $\mathcal{S}$ can be approximated by 1 for all the terms that contribute to the sum. Likewise, defining $\Delta^{\prime}_{S}$ as in (\ref{eqemergedelta1}) and assuming $\Delta_S \lesssim \Delta^{\prime}_{S}$, we get 
\begin{multline}
\text{Tr} \left\langle \rho^{2}_{L} \right\rangle_{\mathrm{EE}} \\ 
=\left(\sum_{i^{}_{1_{S}},i^{}_{1_{L}}} \left| c^{}_{ i^{}_{1_{S}}i^{}_{1_{L}}}\right|^2 Z_{i^{}_{1_L}}\right)\left(\sum_{i^{}_{2_{S}},i^{}_{2_{L}}} \left| c^{}_{ i^{}_{2_{S}}i^{}_{2_{L}}}\right|^2 \right)+\left(\sum_{i^{}_{1_{S}},i^{}_{1_{L}}} \left| c^{}_{ i^{}_{1_{S}}i^{}_{1_{L}}}\right|^2 Z_{i^{}_{1_L}}\right)\left(\sum_{i^{}_{2_{S}},i^{}_{2_{L}}} \left| c^{}_{ i^{}_{2_{S}}i^{}_{2_{L}}}\right|^2 Z_{i^{}_{2_S}}\right)\\
\hspace{8.9cm}+\left(\sum_{i^{}_{1_{S}},i^{}_{1_{L}}} \left| c^{}_{ i^{}_{1_{S}}i^{}_{1_{L}}}\right|^2 \right)\left(\sum_{i^{}_{2_{S}},i^{}_{2_{L}}} \left| c^{}_{ i^{}_{2_{S}}i^{}_{2_{L}}}\right|^2Z_{i^{}_{2_S}} \right)\\ \label{traceeq2} 
\end{multline}
We will show in appendix \ref{appcorrectdelta} that as long as $\frac{\Delta_L}{\bar{U}_L},\frac{\Delta_S}{\bar{U}_S} \ll 1$\footnote{Note that this is precisely what one would expect when the system is in approximate equilibrium in each box, as expected during an epoch.}, we can relax the conditions $\Delta_L \lesssim \Delta^{\prime}_{L}$ and $\Delta_S \lesssim \Delta^{\prime}_{S}$ and still obtain the same eventual entanglement entropy formulas. But the scenario where $\Delta_{L,S} \lesssim \Delta'_{L,S}$ happens to be simple, so we have chosen to emphasize that here in the main body of the paper. It will be interesting to understand what is the relevant physics that controls the relative sizes of $\Delta$ and $\Delta'$, but we will not pursue it here. 

From an analogous calculation, we can also see that
\bea
\begin{aligned}
\text{Tr} \left\langle \rho_{L} \right\rangle_{\mathrm{EE}} =\sum_{i^{}_{1_{S}},i^{}_{1_{L}}} \left| c^{}_{ i^{}_{1_{S}}i^{}_{1_{L}}}\right|^2 
\end{aligned}
\eea
Therefore, we can define the normalized\footnote{Alternatively we could have started with a normalized version of the state \eqref{eqfactor}.} purity of the larger box as follows:
\bea
\begin{aligned}
&\frac{\text{Tr} \left\langle \rho^{2}_{L} \right\rangle_{\mathrm{EE}}}{\left(\text{Tr} \left\langle \rho_{L} \right\rangle_{\mathrm{EE}}\right)^2} \\
&=\frac{\sum_{i^{}_{1_{S}},i^{}_{1_{L}}} \left| c^{}_{ i^{}_{1_{S}}i^{}_{1_{L}}}\right|^2 Z_{i^{}_{1_L}}}{\sum_{i^{}_{1_{S}},i^{}_{1_{L}}} \left| c^{}_{ i^{}_{1_{S}}i^{}_{1_{L}}}\right|^2 }+\left(\frac{\sum_{i^{}_{1_{S}},i^{}_{1_{L}}} \left|c^{}_{ i^{}_{1_{S}}i^{}_{1_{L}}}\right|^2 Z_{i^{}_{1_L}}}{\sum_{i^{}_{1_{S}},i^{}_{1_{L}}} \left| c^{}_{ i^{}_{1_{S}}i^{}_{1_{L}}}\right|^2 }\right)\left(\frac{\sum_{i^{}_{2_{S}},i^{}_{2_{L}}} \left| c^{}_{ i^{}_{2_{S}}i^{}_{2_{L}}}\right|^2Z_{i^{}_{2_S}} }{\sum_{i^{}_{2_{S}},i^{}_{2_{L}}} \left| c^{}_{ i^{}_{2_{S}}i^{}_{2_{L}}}\right|^2}\right)\\
&\hspace{8cm}+\frac{\sum_{i^{}_{2_{S}},i^{}_{2_{L}}} \left| c^{}_{ i^{}_{2_{S}}i^{}_{2_{L}}}\right|^2Z_{i^{}_{2_S}} }{\sum_{i^{}_{2_{S}},i^{}_{2_{L}}} \left| c^{}_{ i^{}_{2_{S}}i^{}_{2_{L}}}\right|^2}
\end{aligned} \label{eqpurityfinal00}
\eea
This expression can be further simplified to
\bea
\begin{aligned}
\frac{\text{Tr} \left\langle \rho^{2}_{L} \right\rangle_{\mathrm{EE}}}{\left(\text{Tr} \left\langle \rho_{L} \right\rangle_{\mathrm{EE}}\right)^2} &= \text{Tr}(\tilde{\rho}_{L}\mathcal{I}_{L}) +\text{Tr}(\tilde{\rho}_{L}\mathcal{I}_{L})\text{Tr}(\tilde{\rho}_{S}\mathcal{I}_{S}) +\text{Tr}(\tilde{\rho}_{S}\mathcal{I}_{S}),\label{eqpurityfinal}
\end{aligned}
\eea
where we have defined 
\bea
\mathcal{I}_{L} = \sum_{i^{}_{L}} Z_{i^{}_{L}} \ | \phi_{i^{}_{L}}\ket\bra \phi_{i^{}_{L}}| \ \ \text{and} \ \ \mathcal{I}_{S} = \sum_{i^{}_{S}} \ Z_{i^{}_{S}} | \psi_{i^{}_{S}}\ket\bra  \psi_{i^{}_{S}}|
\eea
and the normalized reduced density matrices of the boxes as follows:
\bea
\tilde{\rho}_{L} = \frac{1}{\sum_{i^{}_{1_{S}},i^{}_{1_{L}}} \left| c^{}_{ i^{}_{1_{S}}i^{}_{1_{L}}}\right|^2} \sum_{i^{}_{L},i^{\prime}_{L},i^{}_{S}} c^{}_{i^{}_{S}i^{}_{L}}c^{*}_{i^{}_{S}i^{\prime}_{L}} \ | \phi_{i^{}_{L}}\ket\bra \phi_{i^{\prime}_{L}}|
\eea
\bea
\tilde{\rho}_{S} = \frac{1}{\sum_{i^{}_{1_{S}},i^{}_{1_{L}}} \left| c^{}_{ i^{}_{1_{S}}i^{}_{1_{L}}}\right|^2} \sum_{i^{}_{S},i^{\prime}_{S},i^{}_{L}} c^{}_{i^{}_{S}i^{}_{L}}c^{*}_{i^{\prime}_{S}i_{L}} \ | \psi_{i^{}_{S}}\ket\bra \psi_{i^{\prime}_{S}}|
\eea
Here $| \phi_{i^{}_{L}}\ket$ and $| \psi_{i^{}_{S}}\ket$ are eigenstates of the larger and smaller box respectively.
\section{Non-Crossing Partitions and the $n$-th Renyi Entropy}
\begin{figure}
\centering
\begin{subfigure}{.5\textwidth}
  \centering
  \includegraphics[width=.65\linewidth]{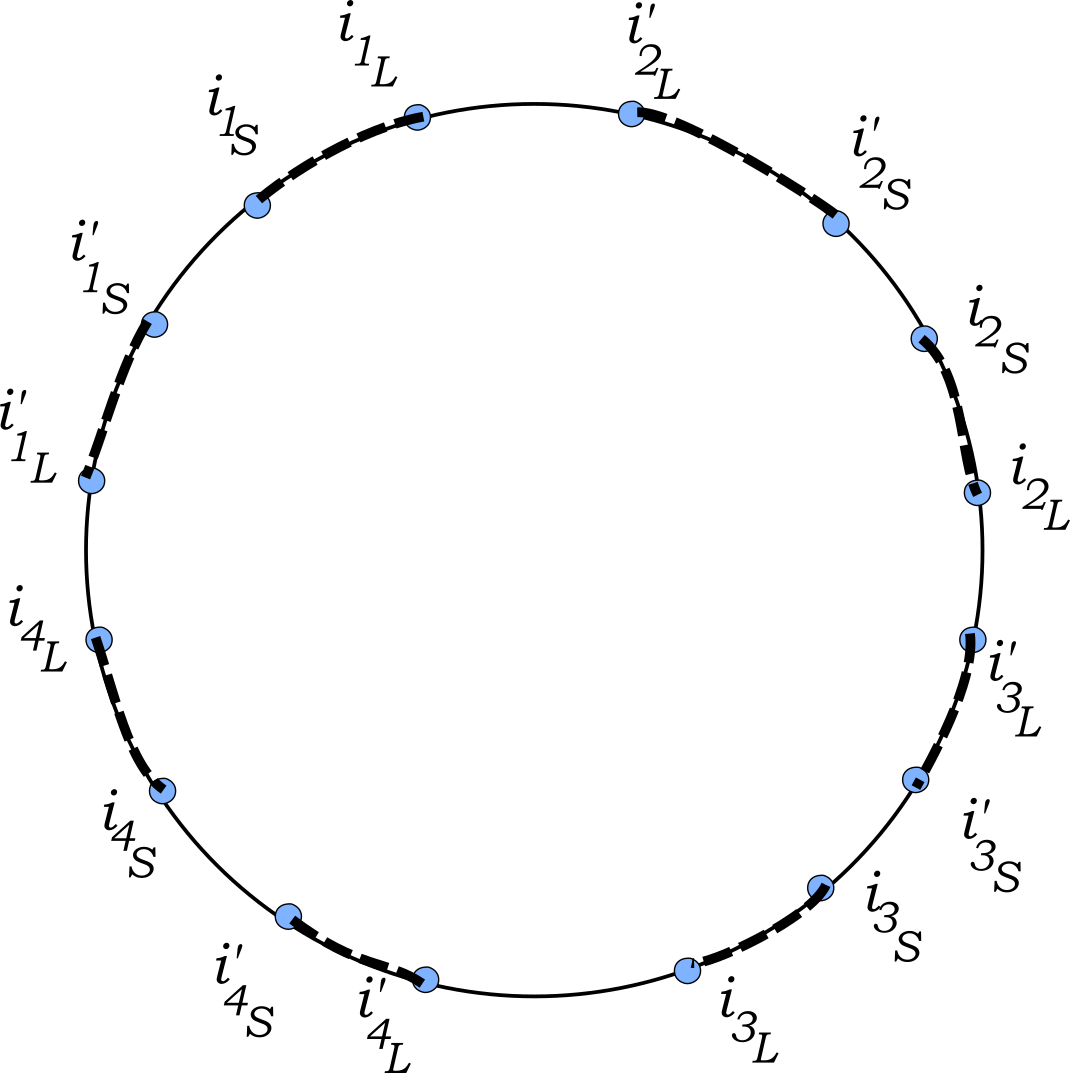}
  \caption{}
\end{subfigure}%
\begin{subfigure}{.5\textwidth}
  \centering
  \includegraphics[width=.65\linewidth]{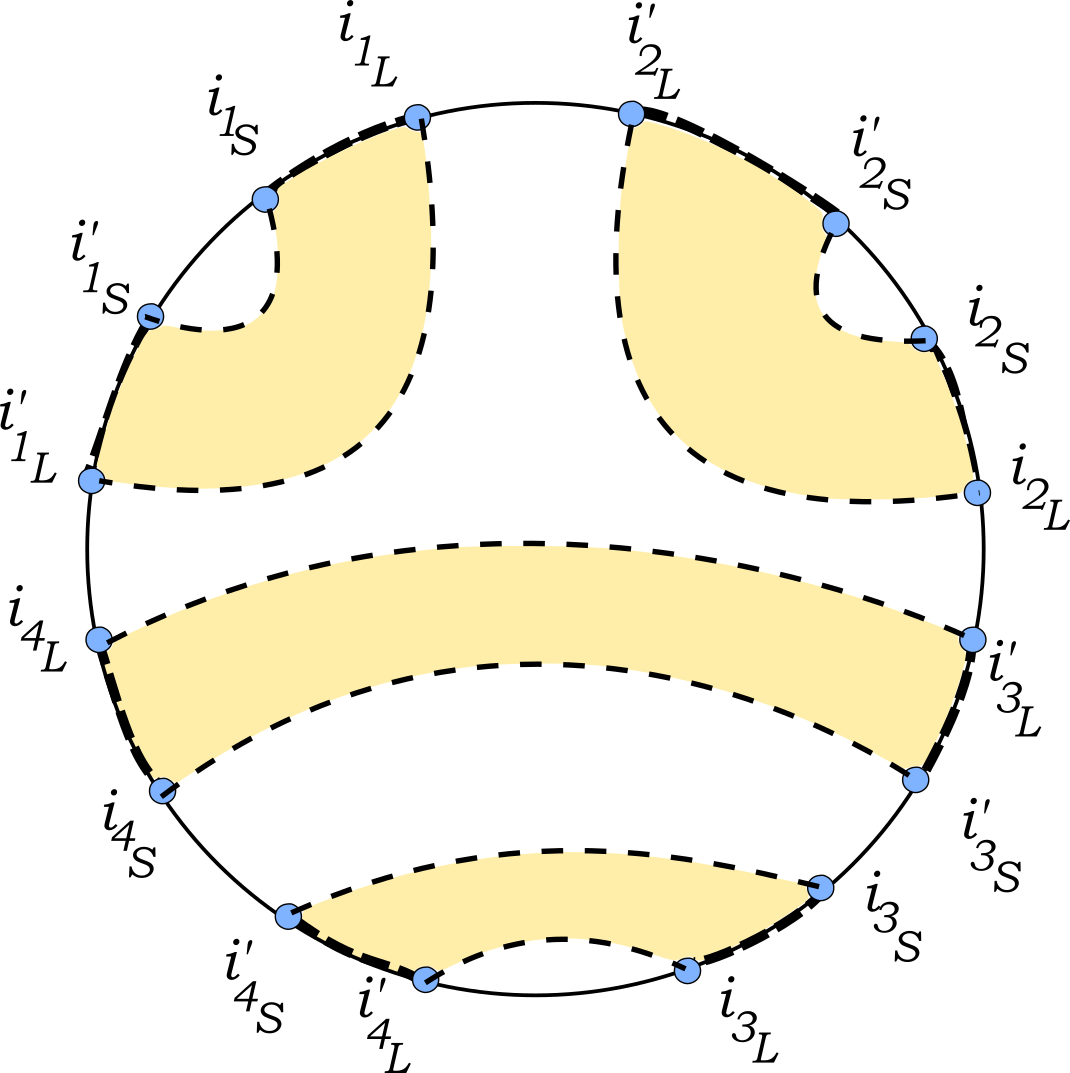}
  \caption{}
\end{subfigure}
\caption{(a) Consider the terms in the computation of the $4$th Renyi entropy. We start off by distributing the indices on a circle. The pair of indices belonging to the same copy of the system are connected by a thick dashed line through the boundary. (b) Connecting various pairs of such indices with each other will gives us a particular contraction.}\label{fig1}
\end{figure}
\noindent Now let us look at the computation of the $n$-th Renyi entropy of the larger box. It is straightforward to calculate $\text{Tr} \left( \rho^{n}_{L} \right)$ by following the steps in section \ref{puritysec}. The resulting expression will contain a product of $2n$ eigenfunctions, as in \eqref{eqrho2}. However, to evaluate this quantity in the eigenstate ensemble, we will have to perform all the possible pairwise contractions of these $2n$ eigenfunctions and then do integrals over the resulting expressions. This can turn out to be quite tedious as the number of possible contractions go as $\frac{(2n )!}{n ! 2^n}$ for a generic $n$, see eg. \cite{Catalan}. Fortunately, we can directly calculate the end result by resorting to a diagrammatic approach.

We start off by distributing all the $4n$ indices present in the higher dimensional analogue of \eqref{traceeq1} on a circle as in fig \ref{fig1}(a). The pair of indices corresponding to each copy of the system are connected by a dotted line through the \textit{boundary} of the circle. Note that these pairs of indices are placed in such a way that the indices of various copies of the smaller (larger) box are adjacent to each other. Now let us connect one such pair of indices to another though the interior of the circle using dashed lines. While making the connection, we make sure that an index corresponding to the smaller (larger) box is connected only to another smaller (larger) box index. Doing this for all the pairs on the circle, we will get a diagram that corresponds to a particular pairwise contraction of all the $2n$ eigenfunctions (Refer \ref{fig1}(b)).

\begin{figure}
\centering
\begin{subfigure}{.5\textwidth}
  \centering
  \includegraphics[width=.65\linewidth]{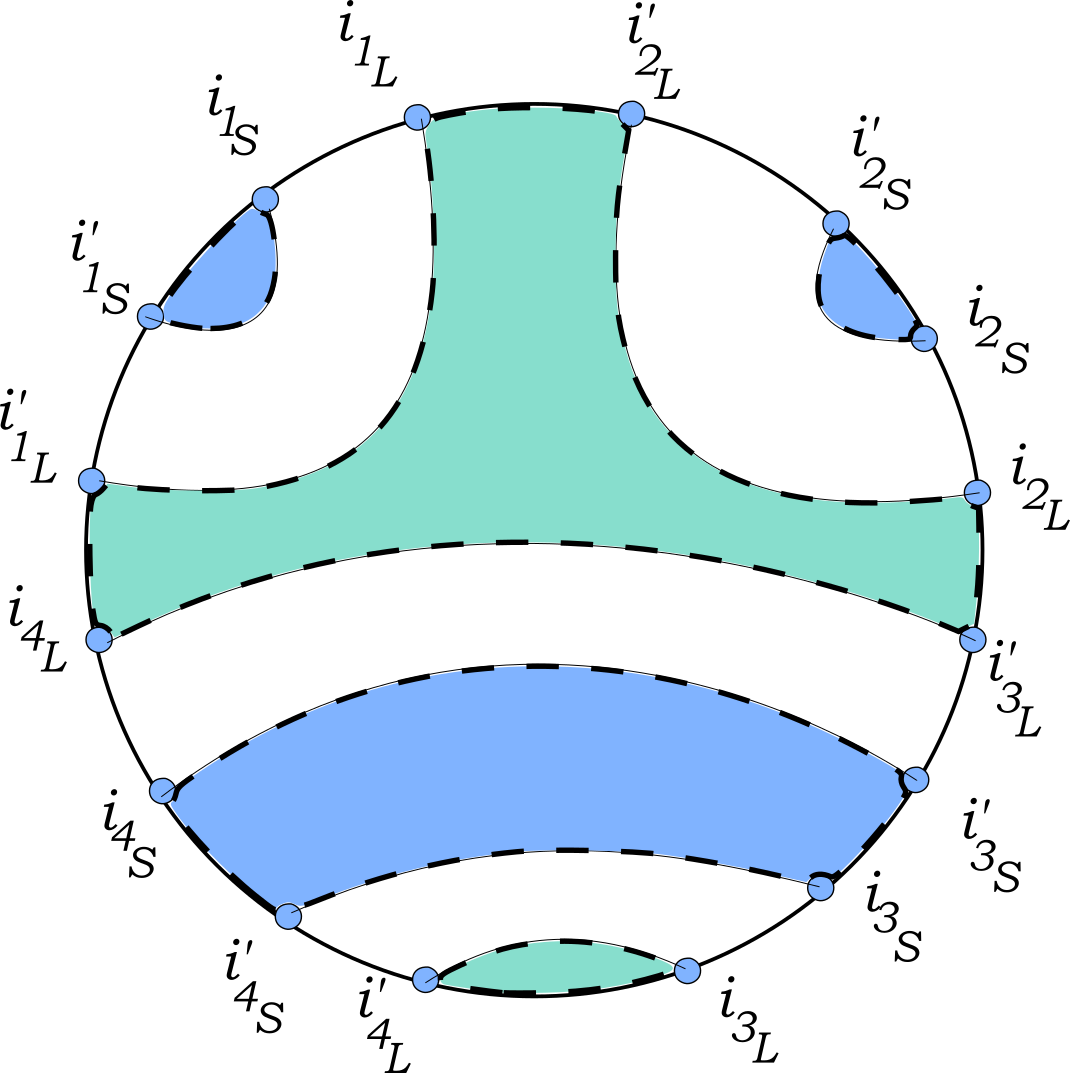}
  \caption{}
\end{subfigure}%
\begin{subfigure}{.5\textwidth}
  \centering
  \includegraphics[width=.65\linewidth]{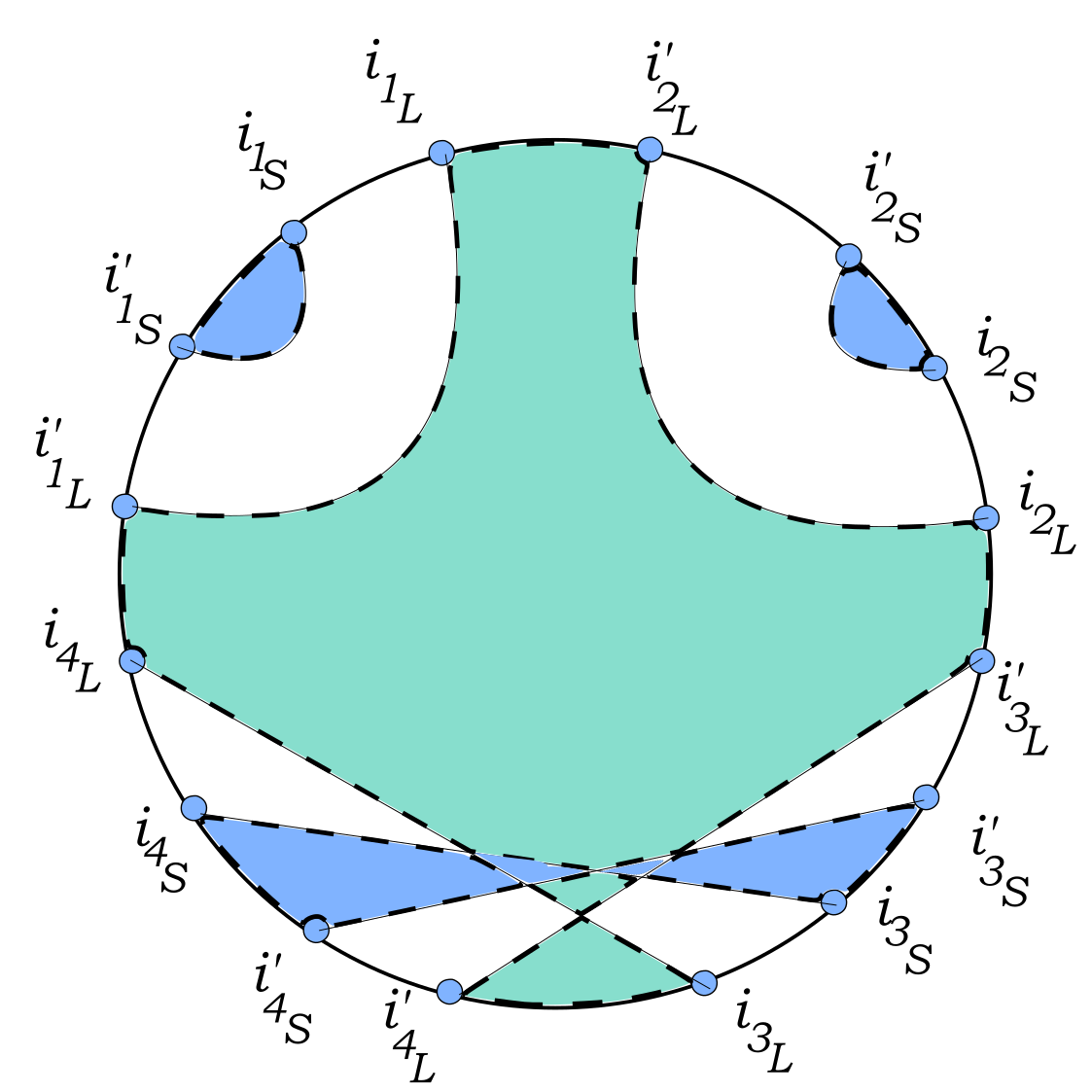}
  \caption{}
\end{subfigure}
\caption{The figure highlights the $m$-connected regions of two diagrams. Let us look at figure (a). This is the same diagram as figure in \ref{fig1}(b). We can see how the interior dashed lines partition the disk into various sub-regions. In this figure, there are two 1-connected regions and one 2-connected region corresponding to the indices of the smaller box (marked in blue) and there are one 1-connected and one 3-connected regions corresponding to the larger box (marked in green). Therefore, this diagram will have a factor of $(\text{Tr}(\tilde{\rho}_{L}\mathcal{I}_{L}))^2\text{Tr}(\tilde{\rho}_{S}\mathcal{I}_{S})$. Similarly, the figure (b) will have a factor of $(\text{Tr}(\tilde{\rho}_{L}\mathcal{I}_{L}))^3\text{Tr}(\tilde{\rho}_{S}\mathcal{I}_{S})$. We can see that this crossing diagram will be sub-leading to figure (a) at every epoch. }\label{fig2}
\end{figure}

Now let us compute the value of each of these diagrams. It is useful to introduce some terminology before we proceed. The dashed lines partition the interior of the circle into various sub-regions (Refer figure \ref{fig2}). Let us call such a sub-region an $m$-connected region if there are $m$ pairs of indices on the boundary of the region. Depending on the box to which the boundary indices belong to, we can attribute each $m$-connected region to the smaller or larger box. For example, in fig \ref{fig2}(a), there are two 1-connected and one 2-connected regions belonging to the smaller box (These regions are marked in blue). 

In terms of these regions, we can assign a value to each diagram by using \eqref{eqZcondition1} and the structure of the contractions. \textit{For every $m$-connected region, we should introduce a factor of $(\text{Tr}(\tilde{\rho}_{S}\mathcal{I}_{S}))^{m-1}$  or $\left(\text{Tr}(\tilde{\rho}_{L}\mathcal{I}_{L})\right)^{m-1}$, depending on which box the region belongs to}. Summing over the value of each of these diagrams will give us the $n$-th Renyi entropy of the larger box. 

If any interior dashed line of a diagram intersect another, then we will refer to these diagram as a crossing diagram. As $\text{Tr}(\tilde{\rho}_{L}\mathcal{I}_{L}),\text{Tr}(\tilde{\rho}_{S}\mathcal{I}_{S}) \ll 1$ at every epoch, it is very easy to see that all the crossing diagrams are sub-leading to the non-crossing diagrams (Refer figure \ref{fig2}(a) and (b) for an example). Therefore, it suffices to add the dominant non-crossing diagrams to get the $n$-th Renyi entropy. This makes the computation easier as the number of non-crossing partitions, called the Catalan number (see eg., \cite{Rodica}), is much smaller than the number of pairwise contractions. We can see that there is a similar contraction structure as well as leading order behavior in \cite{Liu}. This close resemblance has to do with the fact that the ``equilibrium approximation'' in \cite{Liu} is equivalent to a time averaging when the system has reached an approximate (local) equilibrium.

\begin{figure}
\centering
\begin{subfigure}{.5\textwidth}
  \centering
  \includegraphics[width=.65\linewidth]{fig1}
  \caption{}
\end{subfigure}%
\begin{subfigure}{.5\textwidth}
  \centering
  \includegraphics[width=.65\linewidth]{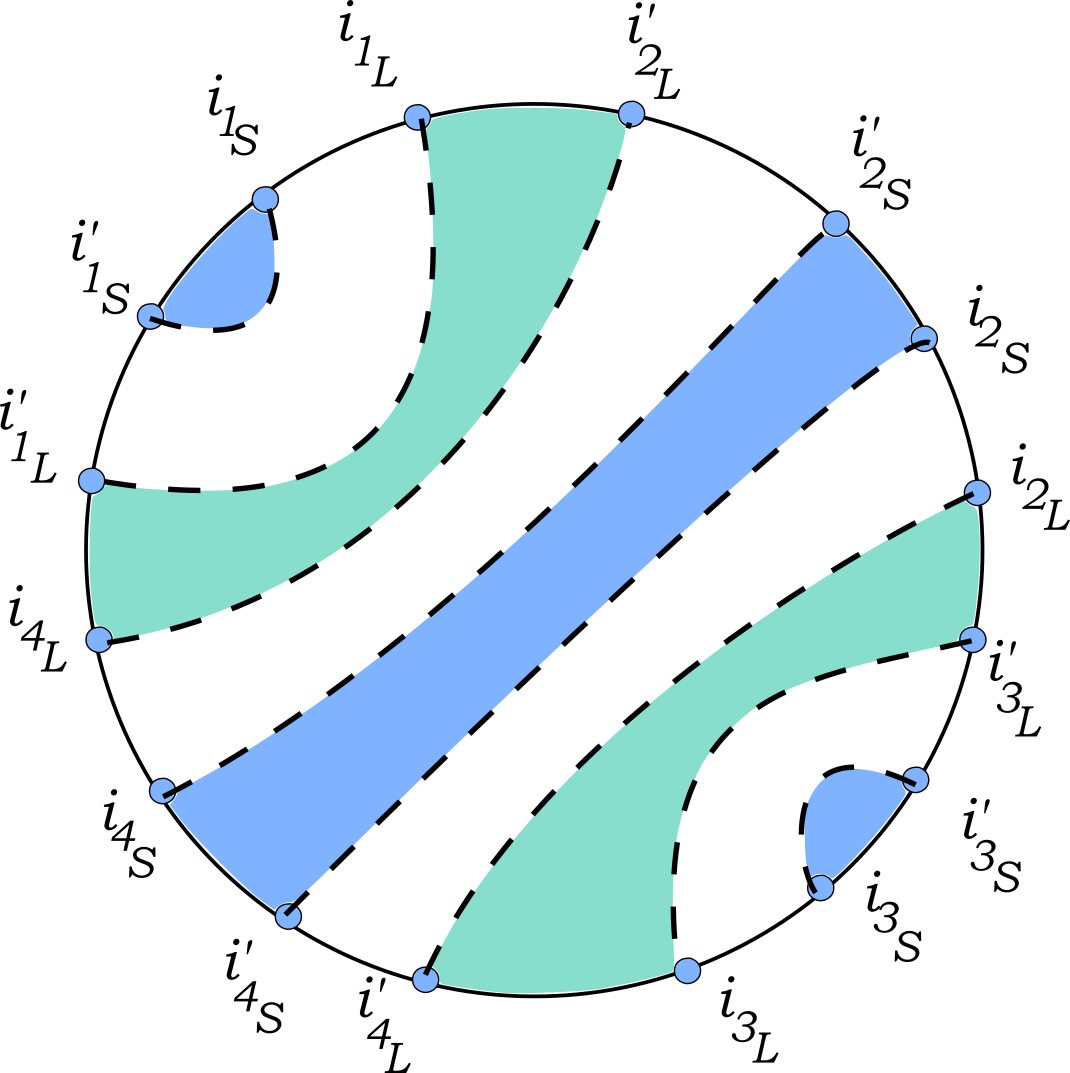}
  \caption{}
\end{subfigure}
\caption{The figure shows two non-crossing diagrams that have the same value. Counting the number of $m$-connected regions, we can see that both the diagram will have a value of $(\text{Tr}(\tilde{\rho}_{L}\mathcal{I}_{L}))^2\text{Tr}(\tilde{\rho}_{S}\mathcal{I}_{S})$. }\label{fig3}
\end{figure}

Now let us write an explicit expression for the $n$-th Renyi entropy by adding the value of all the leading order diagrams. The structure of the contractions results in a large number of degenerate diagrams. Two diagrams can have the same value if one of them can be obtained by permuting of the $m$-connected regions of the other diagram. We can also have a degeneracy when the diagrams have different $m$-connected regions but the powers of $(\text{Tr}(\tilde{\rho}_{L}\mathcal{I}_{L}))$ and $\text{Tr}(\tilde{\rho}_{S}\mathcal{I}_{S})$ add up to the same number (Refer figure \ref{fig3} for an example). 

To take care of these issues, let us first characterize each diagram by the $m$-connected regions of the larger box\footnote{The pairwise contraction structure automatically fixes the $m$-connected regions of the smaller box in the terms of the $m$-connected regions of the larger box. Therefore, it suffices to use either one of the regions to characterize the diagram.}. We can represent the $m$-connected regions of the larger box by the notation $(1^{m_1}2^{m_2}3^{m_3}...{n}^{m_{n}})$, where the number $m_1$ indicates that the diagram contains $m_1$ 1-connected regions, $m_2$ indicates that the diagram contains $m_2$ 2-connected regions, and so on. 
The total number of non-crossing diagrams with the same sub-region structure $(1^{m_1}2^{m_2}3^{m_3}...{n}^{m_{n}})$ is given by \cite{Rodica}
\bea
\mathcal{N}\left(1^{m_1}2^{m_2}3^{m_3}...{n}^{m_{n}}\right)=\frac{n(n-1) \ldots(n-b+2)}{m_{1} ! \ldots m_{n} !} \ \ \ b>1
\eea
where $b = \sum_{i}m_i$. When $b=1$, $\mathcal{N}=1$.

To account for the second type of degeneracies, let us first look at the partitions of a natural number $m$, that is, we look at the all the possible ways in which $m$ can be written as a sum over positive integers. We can label each partition by the set $\{(j,m_j)\}$, where $j \in \mathbb{Z}^{+}$ and $m_j$ corresponds to the multiplicity of each $j$. Therefore, by definition, we have
\bea
j \geq 1 \ \ \ \text{and} \ \ \ \sum_{j}jm_j = m
\eea
Let us denote $P(m)$ to be the set of all such partitions of $m$. Using these definitions, we can write down the leading order contribution to the value of the following expression as:
\bea
\frac{\text{Tr} \left\langle \rho^{n}_{L} \right\rangle_{\mathrm{EE}}}{\left(\text{Tr} \left\langle \rho_{L} \right\rangle_{\mathrm{EE}}\right)^n} =\sum_{k=0}^{n-1}\left(\sum_{\{(j,m_j)\}\in P(k)}\mathcal{N}\left(1^{n-r}\{(j+1)^{m_j}\}\right) \right)(\text{Tr}(\tilde{\rho}_{L}\mathcal{I}_{L}))^{k}(\text{Tr}(\tilde{\rho}_{S}\mathcal{I}_{S}))^{n-k-1} \label{eqentanglement}
\eea
where $r = \sum_{j}(j+1)m_j$. We have used the notation $\left(1^{n-r}\{(j+1)^{m_j}\}\right)$ to represent the non-crossing diagram consisting of $(n-r)$ 1-connected regions and $m_j$ $(j+1)$-connected regions, for all $(j,m_j) \in\{(j,m_j)\}$. When $k=0$ and $k=(n-1)$, we can see that factor in the parenthesis turn out to be 1.

Using the above expression, we can calculate (the leading order contribution to) the averaged $n$-th Renyi entropy,
\bea
\left\langle S_{n}( \rho_{L}) \right\rangle_{\mathrm{EE}} = \frac{1}{1-n}\log\left[\frac{\text{Tr} \left\langle \rho^{n}_{L} \right\rangle_{\mathrm{EE}}}{\left(\text{Tr} \left\langle \rho_{L} \right\rangle_{\mathrm{EE}}\right)^n}\right]. \label{eqentanglement1}
\eea

\section{The Semi-Classical Page Curve}
\noindent To make explicit statements about the behavior of the entanglement entropy, let us calculate the quantities on the RHS of \eqref{eqentanglement}. When $\frac{\Delta_L}{\bar{U}_L},\frac{\Delta_S}{\bar{U}_S} \ll 1$, we can see from appendix \ref{correctionappendix} that
\bea
\text{Tr}(\tilde{\rho}_{L}\mathcal{I}_{L}) \simeq \left(\frac{L^{\prime}}{h}\right)^{-3N_L}\frac{\Gamma(3N_L/ 2) (2 m \bar{U}_L)}{(2 \pi m \bar{U}_L)^{3N_L / 2}} \label{eqaveng1}
\eea
and
\bea
\text{Tr}(\tilde{\rho}_{S}\mathcal{I}_{S}) \simeq \left(\frac{L}{h}\right)^{-3N_S}\frac{\Gamma(3N_S/ 2) (2 m \bar{U}_S)}{(2 \pi m \bar{U}_S)^{3N_S / 2}} \label{eqaveng2}
\eea
To understand the behavior of the entanglement entropy, let us look at early and late times separately. 

When we make plots, we will assume that the average energy per particle is roughly constant. It is possible to relax this assumption somewhat, while retaining the shape of the Page curve, but we will not explore it here since it is quite reasonable as a physical assumption in a closed system of large number of particles  \cite{Srednicki1}.

\subsection{Early time behavior}
\noindent At early times, the larger box will have very small number of particles compared to the smaller box. Therefore, $\text{Tr}(\rho_{L}\mathcal{I}_{L}) \gg \text{Tr}(\rho_{S}\mathcal{I}_{S})$. The $n$-th Renyi entropy will be dominated by the $k=n-1$ term in \eqref{eqentanglement}. The von Neumann entropy of the reduced density matrix can be then calculated as
\bea
\begin{aligned}
\left \langle S (\rho_{L}) \right \rangle_{\mathrm{EE}} &= \lim_{n \to  1} \left \langle S_{n} (\rho_{L}) \right \rangle_{\mathrm{EE}}  \\
&= \lim_{n \to  1} \frac{1}{1-n}\log\left[\frac{\text{Tr} \left\langle \rho^{n}_{L} \right\rangle_{\mathrm{EE}}}{\left(\text{Tr} \left\langle \rho_{L} \right\rangle_{\mathrm{EE}}\right)^n}\right] \\
&= -\log{\text{Tr}(\tilde{\rho}_{L}\mathcal{I}_{L}) } = -\log{\left[ \left(\frac{L^{\prime}}{h}\right)^{-3N_L}\frac{\Gamma(3N_L/ 2) (2 m \bar{U}_L )}{(2 \pi m \bar{U}_{L})^{3N_L / 2}}\right]} 
\end{aligned}\label{entropytranseq}
\eea
For large $N$, $\log{\Gamma(3N/ 2)} = (3N/2-1)\log{3N/2}-3N/2$. This gives us
\bea
\begin{aligned}
\left \langle S (\rho_{L}) \right \rangle_{\mathrm{EE}} &= N_L\log{\left[\frac{V_L (2m\bar{U}_{L})^{3/2}}{h^{3}}\right]} + N_L\left[\log{\left(\frac{2 \pi}{3N_L}\right)^{3/2}}+\frac{3}{2}\right] +O\left(\frac{\log{N_L}}{N_L}\right)\\
& \simeq  N_L\left\{\log{\left[V_L\left(\frac{4\pi m\bar{U}_{L}}{3 h^2 N_L}\right)^{3/2}\right]} +\frac{3}{2}\right\}
\end{aligned} \label{finalentropyeq1}
\eea
where $V_L$ is the volume of the larger box. This is precisely the thermodynamic entropy of the larger box as a function of the number of particles $N_L$ at a given epoch. As we discussed above, if we assume that the average energy per particle is roughly constant, as in \cite{Srednicki1}, we immediately see that the entanglement entropy will increase with time as $N_L$ increases with time. 

\subsection{Late time behavior} 
\begin{figure}
\centering
\includegraphics[scale=0.6]{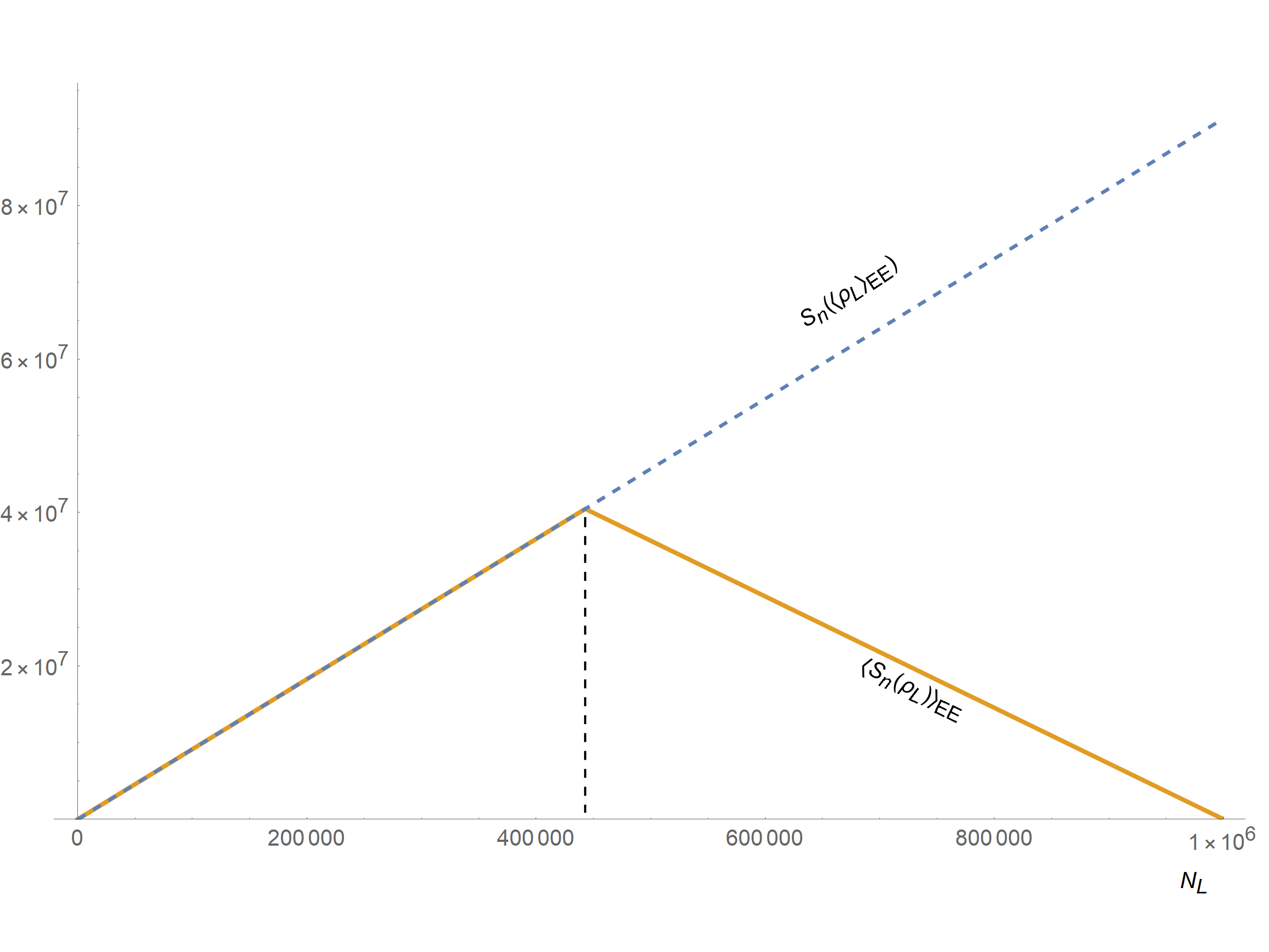}
\caption{The figure shows the plots of $\left \langle S_{n}(\rho_{L}) \right \rangle_{\mathrm{EE}}$ and $S_{n}(\left\langle \rho_{L} \right\rangle_{\mathrm{EE}}) $ (defined in equations \eqref{eqentanglement1} and \eqref{entropyaveragedeq}) as a function of $N_L$ when $n=3$ for a system with $L= 2$ meters, $L^{\prime} = 2^{10}$ meters, $N = 10^6$, and particle mass $m = 1$ amu. We assume that the average energy of the boxes scale linearly with the number of particles as in \cite{Srednicki1}. Therefore, at every epoch, we set $\frac{\bar{U}_{L}}{N_{L}} = \frac{\bar{U}_{S}}{N} = \frac{3}{2} k_B T$ and we choose $T= 300$ K. As more and more particles leak out into the larger box, we can see that the average entanglement entropy (denoted by the orange line) increases with time initially and then reaches a maximum at the Page time (indicated by the dotted vertical line). After the Page time, the entanglement entropy drops and saturates to the equilibrium value. As this value is much smaller than the maximum value at Page time, the entanglement entropy plot would look like it has dropped down to zero. In contrast to this behavior, the entanglement entropy of the averaged state, indicated by the blue dotted line, keeps on relentlessly increasing with time (until it saturates at its final value).}
\label{fig4}
\end{figure}
\noindent With the passage of time, more and more particles start moving into the larger box. This results in an increase in the value of $\text{Tr}(\tilde{\rho}_{S}\mathcal{I}_{S})$ and decrease in $\text{Tr}(\tilde{\rho}_{L}\mathcal{I}_{L})$. Therefore, the entanglement entropy at late times will be behave differently from that of the previous section. Depending upon the relative size of the larger box, there are two types of late time behaviors. 

To understand these behaviors, let us first look at very late times. The system will reach global equilibrium and the net particle exchange between the boxes will drop to zero. This will happen when the particle density in each of the boxes equalize. Let $N_{S_E}$ and $N_{L_E}$ denote the final equilibrium values of the number of particles in the boxes. Therefore, we have the relation
\bea
\frac{N_{L_E}}{L^{\prime}{}^3} = \frac{N_{S_E}}{{L}^3}
\eea
Using 
\bea
N_{L_E}+N_{S_E} = N, \label{eqtotalparticleno}
\eea
we can solve the above equation and we get
\bea
N_{S_E} = \frac{{L}^3N}{L^{\prime}{}^3+{L}^3} \ \ \ \ \ \text{and} \ \ \ \ \ N_{L_E} = \frac{L^{\prime}{}^3N}{L^{\prime}{}^3+{L}^3} \label{numeq}
\eea
Let us define $t_E$ as the time taken for $N_{L_E}$ particles to leak into the larger box.

An intermediate epoch which will be very important to our discussion is characterized by the relation
\bea
\text{Tr}(\tilde{\rho}_{S}\mathcal{I}_{S}) = \text{Tr}(\tilde{\rho}_{L}\mathcal{I}_{L}) \label{eqpagetime}
\eea
We call the time taken to reach this epoch the \textit{Page time} ($t_{P}$) of the system. Let us denoted the number of particles in the smaller and larger boxes at this epoch by $N_{S_P}$ and $N_{L_P}$ respectively. For large $N_{S_P}$ and $N_{L_P}$, we can solve \eqref{eqpagetime} and \eqref{eqtotalparticleno} to get
\bea
N_{L_P} = \frac{\left\{\log{\left[\frac{L^{3}}{\bar{\lambda}_{S}}\right]} +\frac{3}{2}\right\}N}{\log{\left[\frac{L^{3}}{\bar{\lambda}_{S}}\right]} +\log{\left[\frac{{L^{\prime}}^{3}}{\bar{\lambda}_{L}}\right]} +3} \ \ \ \ \ \text{and} \ \ \ \ \ N_{S_P} = \frac{\left\{\log{\left[\frac{{L^{\prime}}^{3}}{\bar{\lambda}_{L}}\right]} +\frac{3}{2}\right\}N}{\log{\left[\frac{L^{3}}{\bar{\lambda}_{S}}\right]} +\log{\left[\frac{{L^{\prime}}^{3}}{\bar{\lambda}_{L}}\right]} +3} \label{peq}
\eea
where $\bar{\lambda}_{S}$ and $\bar{\lambda}_{L}$ are the average thermal wavelength of the boxes. In obtaning the above form of the solution, it is convenient to take the log of \eqref{eqpagetime} and remember the approximations we have made use of in \eqref{finalentropyeq1}. A key useful fact is again that the $\bar U/N_{box}$ for each of the boxes is approximately $N_{box}$-independent, and therefore the equations reduce effectively to linear equations in particle numbers. These are trivially solved, and the result is what is quoted above. As the number of particles in each of the boxes remain a constant after $t_E$, the system will reach the Page time if and only if $N_{L_P} \lesssim N_{L_E}$. When the larger box is sufficiently bigger than the smaller box, we can see that this condition will always be satisfied. This is immediate upon comparing \eqref{numeq} and \eqref{peq}.

Now let us look at the late time behavior of the entanglement entropy when $L^{\prime} \gg L$. As $N_{L_P} \lesssim N_{L_E}$, the system would reach the Page time before it reaches global equilibrium. For any $t\lesssim t_{P}$, $\text{Tr}(\tilde{\rho}_{L}\mathcal{I}_{L}) \gg \text{Tr}(\tilde{\rho}_{S}\mathcal{I}_{S})$. Therefore, the entanglement entropy will have the same trend as the early time behavior. However, when $t\gtrsim t_{P}$, $\text{Tr}(\tilde{\rho}_{L}\mathcal{I}_{L}) \ll \text{Tr}(\tilde{\rho}_{S}\mathcal{I}_{S})$. This would mean that $k=0$ term in the equation \eqref{eqentanglement} will dominate the sum. Therefore, we have
\bea
\begin{aligned}
\left \langle S (\rho_{L}) \right \rangle_{\mathrm{EE}} &= \lim_{n \to  1} \left \langle S_{n} (\rho_{L}) \right \rangle_{\mathrm{EE}}  \\
&= \lim_{n \to  1} \frac{1}{1-n}\log\left[\frac{\text{Tr} \left\langle \rho^{n}_{L} \right\rangle_{\mathrm{EE}}}{\left(\text{Tr} \left\langle \rho_{L} \right\rangle_{\mathrm{EE}}\right)^n}\right] \\
&=-\log{\left[ \left(\frac{L}{h}\right)^{-3N_S}\frac{\Gamma(3N_S/ 2) (2 m \bar{U}_{S})}{(2 \pi m\bar{U}_{S})^{3N_S / 2}}\right]}\\
& \simeq N_S\left\{\log{\left[V_S\left(\frac{4\pi m\bar{U}_{S}}{3 h^2 N_S}\right)^{3/2}\right]} +\frac{3}{2}\right\} \ \ \ \ \ \ \ t\gtrsim t_{P}
\end{aligned} \label{finalentropyeq2}
\eea
The resulting equation is the thermodynamic entropy of the \textit{smaller box}. For $t \gtrsim t_E$, the entanglement entropy will saturate to the value
\bea
\begin{aligned}
\left \langle S (\rho_{L}) \right \rangle_{\mathrm{EE}} &=  N_{S_E}\left\{\log{\left[V_S\left(\frac{4\pi m\bar{U}_{S}}{3 h^2 N_{S_E}}\right)^{3/2}\right]} +\frac{3}{2}\right\}.
\end{aligned} \label{eqentropysaturated}
\eea
If we assume that at each epoch we have $N_L+N_S = N$, then we can plot the entanglement entropy as a function of $N_L$ or $N_S$. We can see from figure \ref{fig2} that the graph increases at early times and the reaches a maximum at the Page time. The graph then decreases and saturates to \eqref{eqentropysaturated}.

Now let us look at the case where the sizes of the boxes are comparable to each other. Depending upon the thermal wavelength of the boxes, we can have either $N_{L_E} \lesssim N_{L_P}$ or $N_{L_E} \gtrsim N_{L_P}$. The behavior of the entanglement entropy will be identical to the discussion in the previous paragraph when $N_{L_E} \lesssim N_{L_P}$. Let us look at the other case in some detail. The system would reach global equilibrium before the Page time is reached. This would mean that the entanglement entropy will keep on increasing and then saturate to the value at $t_E$. Therefore, we have
\bea
\begin{aligned}
\left \langle S (\rho_{L}) \right \rangle_{\mathrm{EE}} = N_{L_E}\left\{\log{\left[V_{L}\left(\frac{4\pi m\bar{U}_{L}}{3 h^2 N_{L_E}}\right)^{3/2}\right]} +\frac{3}{2}\right\} \ \ \ \ \ \ \ \ \  t \gtrsim t_E
\end{aligned} \label{eqentropysaturated1}
\eea
We mention this only for completeness, our primary interest is in the previous scenario.

\section{An Information ``Paradox"}

We have managed to reproduce the Page curve by doing an ensemble averaged computation of the entanglement entropy at each epoch. In this section, we will instead compute the entanglement entropy of the ensemble averaged state at each epoch. We will find that this leads to an information ``paradox". We put the word in quotes because a genuine information paradox is tied to the existence of horizons, and also because here we know why the paradox is appearing.

Let us start by evaluating the reduced density matrix in the eigenstate ensemble:
\bea
 \left\langle \rho_{L} (\mathbf{y};\mathbf{y}^{\prime}) \right\rangle_{\mathrm{EE}} = \int_{D} dx \sum_{\alpha,{\alpha^{\prime}}} d^{}_{\alpha}d^{*}_{ {\alpha^{\prime}}} \ \left\langle\Psi_\alpha(\mathbf{x},\mathbf{y})  \Psi^{*}_{\alpha^{\prime}}(\mathbf{x},\mathbf{y}^{\prime}) \right\rangle_{\mathrm{EE}}
\eea
Squaring the matrix, we get
\bea
\begin{aligned}
 \left\langle \rho_{L}(\mathbf{y};\mathbf{y}^{\prime})  \right\rangle_{\mathrm{EE}}^{2}  &= \int_{D} dy^{\prime \prime}  \  \left\langle \rho_{L}  (\mathbf{y};\mathbf{y}^{\prime \prime})\right\rangle_{\mathrm{EE}} \left\langle \rho_{L}  (\mathbf{y}^{\prime \prime};\mathbf{y}^{\prime}) \right\rangle_{\mathrm{EE}} \\
&= \int_{D} dy^{\prime \prime} \int_{D} dx dx^{\prime}\sum_{\alpha^{}_{1},{\alpha^{\prime}_{1}},\alpha^{}_{2},{\alpha^{\prime}_{2}}} d^{}_{\alpha^{}_{1}}d^{*}_{ {\alpha^{\prime}_{1}}}d^{}_{{\alpha^{}_{2}}}d^{*}_{ {\alpha^{\prime}_{2}}}\\
& \hspace{4cm} \left\langle\Psi_{\alpha^{}_{1}}(\mathbf{x},\mathbf{y})  \Psi^{*}_{\alpha^{\prime}_1}(\mathbf{x},\mathbf{y}^{\prime \prime})\right\rangle_{\mathrm{EE}}\left\langle\Psi_{\alpha^{}_{2}}(\mathbf{x}^{\prime},\mathbf{y}^{\prime \prime})  \Psi^{*}_{\alpha^{\prime}_{2}}(\mathbf{x}^{\prime},\mathbf{y}^{\prime})\right\rangle_{\mathrm{EE}}
\end{aligned}
\eea
Therefore, the purity of the larger box will be given by
\bea
\begin{aligned}
\text{Tr}\left\langle\rho_{L} \right\rangle_{\mathrm{EE}}^{2} & = \int_{D} dy \left\langle\rho_{L}(\mathbf{y};\mathbf{y}) \right\rangle_{\mathrm{EE}}^{2}\\
&=  \int_{D} dy^{\prime \prime}  dy \int_{D} dx dx^{\prime}\sum_{\alpha^{}_{1},{\alpha^{\prime}_{1}},\alpha^{}_{2},{\alpha^{\prime}_{2}}}  d^{}_{\alpha^{}_{1}}d^{*}_{ {\alpha^{\prime}_{1}}}d^{}_{{\alpha^{}_{2}}}d^{*}_{ {\alpha^{\prime}_{2}}} \\
& \hspace{4cm}\left\langle\Psi_{\alpha^{}_{1}}(\mathbf{x},\mathbf{y})  \Psi^{*}_{\alpha^{\prime}_1}(\mathbf{x},\mathbf{y}^{\prime \prime})\right\rangle_{\mathrm{EE}} \left\langle\Psi_{\alpha^{}_{2}}(\mathbf{x}^{\prime},\mathbf{y}^{\prime \prime})  \Psi^{*}_{\alpha^{\prime}_{2}}(\mathbf{x}^{\prime},\mathbf{y})\right\rangle_{\mathrm{EE}}\\
&=  \int_{D} dy^{\prime \prime}  dy \int_{D} dx dx^{\prime} \ \left\langle\Psi(\mathbf{x},\mathbf{y})  \Psi^{*}(\mathbf{x},\mathbf{y}^{\prime \prime})\right\rangle_{\mathrm{EE}} \left\langle\Psi(\mathbf{x}^{\prime},\mathbf{y}^{\prime \prime})  \Psi^{*}(\mathbf{x}^{\prime},\mathbf{y})\right\rangle_{\mathrm{EE}}
\end{aligned}
\eea
Now let us evaluate the two-point functions in the above expression. Using the factorization in \eqref{eqfactor}, we get
\bea
\begin{aligned}
&\text{Tr}\left\langle\rho_{L} \right\rangle_{\mathrm{EE}}^{2} = \int_{D_L} dy^{\prime \prime}  dy \int_{D_S} dx dx^{\prime} \sum_{i^{}_{1_{S}},i^{\prime}_{1_S},i^{}_{1_{L}},i^{\prime}_{1_L},i^{}_{2_{S}},i^{\prime}_{2_S},i^{}_{2_{L}},i^{\prime}_{2_L}}  c^{}_{ i^{}_{1_{S}}i^{}_{1_{L}}}c^{*}_{ i^{\prime}_{1_S}i^{\prime}_{1_L}} c^{}_{i^{}_{2_{S}}i^{}_{2_{L}}}c^{*}_{ i^{\prime}_{2_S}i^{\prime}_{2_L}}\\
&\hspace{3cm} \Bigl{[}\left\langle\psi^{*}_{ i^{\prime}_1}(\mathbf{x})\phi^{*}_{ i^{\prime}_{1}}(\mathbf{y})\psi_{i^{}_{1} }(\mathbf{x})\phi_{ i^{}_{1}}(\mathbf{y}^{\prime \prime})\right\rangle_{\mathrm{EE}} \left\langle\psi^{*}_{i^{\prime}_{2}}(\mathbf{x^{\prime }})\phi^{*}_{ i^{\prime}_{2}}(\mathbf{y}^{\prime \prime})\psi_{i^{}_{2}}(\mathbf{x^{\prime }}) \phi_{ i^{}_{2}}(\mathbf{y})\right\rangle_{\mathrm{EE}} \Bigl{]}
\end{aligned} \label{averagedoutpureq}
\eea
This is precisely the first term of the equation \eqref{traceeq1}. Therefore, we can immediately carryover the calculations in section \ref{puritysec} to get
\bea
\begin{aligned}
\frac{\text{Tr} \left\langle \rho_{L} \right\rangle_{\mathrm{EE}}^{2}}{\left(\text{Tr} \left\langle \rho_{L} \right\rangle_{\mathrm{EE}}\right)^2} &= \text{Tr}(\tilde{\rho}_{L}\mathcal{I}_{L}) )
\end{aligned}
\eea
We can also have
\bea
\begin{aligned}
\frac{\text{Tr} \left\langle \rho_{L} \right\rangle_{\mathrm{EE}}^{n}}{\left(\text{Tr} \left\langle \rho_{L} \right\rangle_{\mathrm{EE}}\right)^n} &= \left(\text{Tr}(\tilde{\rho}_{L}\mathcal{I}_{L}) )\right)^{n-1}
\end{aligned}
\eea
Therefore, the entanglement entropy of the averaged state will be given by
\bea
\begin{aligned}
S(\left\langle \rho_{L} \right\rangle_{\mathrm{EE}}) &= \lim_{n \to  1}S_{n}(\left\langle \rho_{L} \right\rangle_{\mathrm{EE}})\\
&= \lim_{n \to  1} \frac{1}{1-n}\log\left[\frac{\text{Tr} \left\langle \rho_{L} \right\rangle_{\mathrm{EE}}^{n}}{\left(\text{Tr} \left\langle \rho_{L} \right\rangle_{\mathrm{EE}}\right)^n}\right] \\
&= -\log{\text{Tr}(\tilde{\rho}_{L}\mathcal{I}_{L}) } \\
& \simeq  N_L\left\{\log{\left[V_L\left(\frac{4\pi m\bar{U}_{L}}{3 h^2 N_L}\right)^{3/2}\right]} +\frac{3}{2}\right\}
\end{aligned} \label{entropyaveragedeq}
\eea
Under the assumptions of the previous section, we can see that the plot of $S(\left\langle \rho_{L} \right\rangle_{\mathrm{EE}})$ v/s time will keep on increasing and then saturate to the value at $N_L= N_{L_E}$ (see figure \ref{fig4}). Therefore, when the size of the larger is box is sufficiently larger than the smaller box, the late time behavior of $S(\left\langle \rho_{L} \right\rangle_{\mathrm{EE}})$ will be different from that of $\left \langle S (\rho_{L}) \right \rangle_{\mathrm{EE}}$.

We can calculate the purity of the \textit{smaller} box in the eigenstate ensemble by interchanging $\psi \leftrightarrow \phi$, $x \leftrightarrow y$,  $x^{\prime} \leftrightarrow y^{\prime}$ and  $x^{\prime \prime} \leftrightarrow y^{\prime \prime}$ in \eqref{traceeq3} and it turns out to be equal to the purity of the larger box in the eigenstate ensemble. Therefore, we have $\left \langle S (\rho_{L}) \right \rangle_{\mathrm{EE}} = \left \langle S (\rho_{S}) \right \rangle_{\mathrm{EE}}$. This behavior is expected from a unitary theory. However, if we make the same replacements in \eqref{averagedoutpureq}, we will be able to see that $S(\left\langle \rho_{L} \right\rangle_{\mathrm{EE}}) \neq S(\left\langle \rho_{S} \right\rangle_{\mathrm{EE}})$. Therefore, there is an apparent loss of unitarity when we work with the averaged state. 

These results, while simple, are interesting because they provide an explicit mechanism for understanding how the information paradox may emerge in gravity. It suggests that the vacuum one obtains by quantizing fields in the black hole background has features of an $averaged$ state from the perspective of the fundamental theory.  

\section{Semi-Classical Gravity as an Ergodic Effective Theory}

Our calculations in this paper had nothing to do with gravity, horizons or a true information paradox. In fact our primary goal was to illustrate that an ensemble-averaged semi-classical approximation leading to the Page curve is $not$ limited to gravity. But in doing this, we learnt that the ensemble average can arise as a proxy for a time average during each epoch, even in single realizations of a unitary theory. This is interesting because the precise role of the ensemble in the case of gravity has been a bit murky. For one, in 2-d JT gravity there is an explicit ensemble average. But in usual AdS/CFT in higher dimensions, we expect to see black holes in the duals of single copies of the CFT.

In our hard sphere gas, we found that an ensemble average can arise in the ergodic sense during each epoch of local equilibrium. This suggests a similar picture for gravity in higher dimensions. Loosely related ideas have appeared previously in \cite{Sully, Liu}, and our goal here was to find a model that provides a nuts-\&-bolts understanding of the origin of the Page curve. The Page time of the black hole is vastly larger than its scrambling time, and Hawking temperature is a well-defined approximately constant quantity during any epoch of evaporation. This makes it possible that the ensemble average in gravity is a proxy for a time average during each epoch of Hawking radiation. In other words, an explicit average over an ensemble of distinct unitary theories may not be necessary. 

A further observation we made is that the Page version of the Hawking paradox can emerge in our perfectly unitary system, if we did our semi-classical calculation using the ensemble-averaged state. We showed that the entropy increases relentlessly until it saturates at the thermodynamic entropy. This again is a strong suggestion that a similar mechanism may be at work in the gravity system as well -- indeed, a proposal that the Hawking result is a consequence of an ensemble-averaged state was suggested previously in \cite{Bousso}. 

In fact, our calculation in Section 6 demonstrates that the ``state paradox" formulated in \cite{Bousso} can be resolved without an $explicit$ ensemble average over many theories. Let us take a moment to explain this.  We start with the Quantum Extremal Surface \cite{Wall}  formula for the von Neumann entropy, which we write schematically as 
\bea
S_{micro}=\frac{A}{4G} + S_{macro}, \label{QES}
\eea
where the $S_{macro}$ on the right hand side is the entropy of the bulk quantum fields as would have been calculated by Hawking. State paradox arises, if we view the latter as a fine-grained contribution to the full entropy. It was proposed in \cite{Bousso} that the problem can be solved if we view $S_{micro}$ as an ensemble average of the entropy of the state, and $S_{macro}$ as an entropy of the ensemble averaged state. Our calculations in this paper suggest that these ensembles need not be explicit collections of distinct theories like in JT gravity --- they can be ergodic ensembles that stand in for epoch time averages. In particular, the ensembles one thinks of here are not fixed, they are implicitly epoch-dependent. In our example of the hard sphere gas, it is controlled by the number of particles in either box in a given epoch, which fixes the appropriate Berry's ensemble.

A key point in the above discussion is that the $S_{macro}$ on the right hand side of \eqref{QES} is supposed to be computed semi-classically, via a state obtained from quantum field theory in curved space. Even though this object is usually viewed as a fine-grained entropy in the QFT in curved space Hilbert space, it is not clear that it is a fine-grained quantity in the true microscopic degrees of freedom in the holographic CFT Hilbert space. This was emphasized in \cite{CK1} where the $S_{macro}$ was called a ``coarse-grained" entropy. Let us emphasize that this is distinct from some of the uses of the phrase ``coarse-graining" in the literature. What \cite{CK1} emphasized was that the bulk entropy is calculated in a semi-classical bulk state, and not in the truly microscopic CFT state. The precise connection between the two descriptions has never been very clear in AdS/CFT, but observations in this paper suggest that the states in the Hilbert space of a quantum field theory in the black hole background has features of an averaged state. We will elaborate on this preliminary observation, in future work \cite{More}. More generally, if taken at face value, the message of our work is that semi-classical gravity should be viewed as a tool for capturing ergodic averaged gravitational dynamics, for evolution that is in  bulk local equilibrium. Of course, developing this idea further is something that will have to be left for future work. In our hard sphere calculation, the assumption of local equilibrium entered due to the hierarchical timescales that ensured the existence of epochs. See \cite{torr} for another discussion of hierarchical timescales in strongly coupled theories, which may be related to black hole physics.

%Before concluding, let us make one more comment. Entanglement entropy is a fine-grained quantity, so it may seem surprising that an ergodic averaged calculation is able to see it. We believe this is essentially the same puzzle as the one we raised as the first bullet point at the beginning of the Introduction. The answer is also the same -- while the entanglement entropy of a density matrix depends on fine-grained details, it is ultimately just one number. So it is not in principle impossible that a semi-classical calculation is able to see it. Of course, it is an interesting question how this ``UV-independence" emerges and why semi-classical physics is as powerful as it seems to be.

We have been quite specific in our focus in this paper, but let us conclude by emphasizing that the recent ideas on the black hole Page curve\footnote{See \cite{Laddha} for another take on the Page curve in gravity.} may have implications even beyond black holes. See eg. closely related discussions in cosmology \cite{CK2, Malda2, Hartman, Bala, Manu}. We refer the reader to eg. \cite{Kundu, Malvimat, Choudhury, Wang, Ka, Karan} for more recent papers with a fairly thorough list of references.

\section{Acknowledgments}

We thank Jude Pereira for discussions and collaborations.

\appendix
\section{Quantum Chaos in the Hard Sphere Gas}
\label{npointapp}
\noindent Let us consider $N$ hard spheres, of radii $a$, enclosed in a single cubic box of length $L+2a$ as in \cite{Srednicki1}\footnote{Note the technical caveat we made in footnote \ref{4}. For the single box, we can use either language -- normal spheres with box size $L+2a$ or spheres whose center can reach the box walls, with box size $L$. The two problems are mathematically identical.}. We can denote the energy eigenfunctions of the system by $\psi_i(\mathbf{X})$, where $\mathbf{X} = (\mathbf{x}_{1}, \ldots, \mathbf{x}_{N})$ is a $3N$-dimensional vector that labels the position of all the particles of the system. We can define $\psi_i(\mathbf{X})$ on the domain 
\bea
\mathcal{D}=\left\{\mathbf{x}_{1}, \ldots, \mathbf{x}_{N}\left|-\frac{1}{2} L \leq x_{i 1,2,3} \leq+\frac{1}{2} L ;\right| \mathbf{x}_{i}-\mathbf{x}_{j} \mid \geq 2 a\right\}
\eea
satisfying the boundary condition that $\psi_i(\mathbf{X})$ vanishes on $\partial \mathcal{D}$.

The eigenfunctions of the box can be chosen to be real and they take the form \cite{Srednicki1}
\bea
\psi_{\alpha}(\mathbf{X})=\mathcal{N}_{\alpha} \int d^{3 N} P A_{\alpha}(\mathbf{P}) \delta\left(\mathbf{P}^{2}-2 m U_{\alpha}\right) \exp (i \mathbf{P} \cdot \mathbf{X} / \hbar)
\eea
where $A_{\alpha}^{*}(\mathbf{P})=A_{\alpha}(-\mathbf{P})$ and $U_\alpha$ is the energy of the state. Berry's conjecture says that when the energy of $\psi_\alpha(\mathbf{X})$ is sufficiently high, $A_{\alpha}(\mathbf{P})$ acts as if it is a Gaussian random variable with the two-point function 
\bea
\left\langle A_{\alpha}(\mathbf{P}) A_{\beta}\left(\mathbf{P}^{\prime}\right)\right\rangle_{\mathrm{EE}}=\delta_{\alpha \beta} \delta^{3 N}\left(\mathbf{P}+\mathbf{P}^{\prime}\right) / \delta\left(\mathbf{P}^{2}-\mathbf{P}^{\prime 2}\right) \label{atwopoint}
\eea 
Moreover, we can compute the four-point functions in terms of the two-point functions as follows
\bea
\begin{aligned}
\left\langle A_{\alpha}\left(\mathbf{P}_{1}\right) A_{\beta}\left(\mathbf{P}_{2}\right) A_{\gamma}\left(\mathbf{P}_{3}\right) A_{\delta}\left(\mathbf{P}_{4}\right)\right\rangle_{\mathrm{EE}} &=\left\langle A_{\alpha}\left(\mathbf{P}_{1}\right) A_{\beta}\left(\mathbf{P}_{2}\right)\right\rangle_{\mathrm{EE}}\left\langle A_{\gamma}\left(\mathbf{P}_{3}\right) A_{\delta}\left(\mathbf{P}_{4}\right)\right\rangle_{\mathrm{EE}} \\
&+\left\langle A_{\alpha}\left(\mathbf{P}_{1}\right) A_{\gamma}\left(\mathbf{P}_{3}\right)\right\rangle_{\mathrm{EE}}\left\langle A_{\delta}\left(\mathbf{P}_{4}\right) A_{\beta}\left(\mathbf{P}_{2}\right)\right\rangle_{\mathrm{EE}} \\
&+\left\langle A_{\alpha}\left(\mathbf{P}_{1}\right) A_{\delta}\left(\mathbf{P}_{4}\right)\right\rangle_{\mathrm{EE}}\left\langle A_{\beta}\left(\mathbf{P}_{2}\right) A_{\gamma}\left(\mathbf{P}_{3}\right)\right\rangle_{\mathrm{EE}}
\end{aligned}
\eea
One can also write down a similar ``factorization" condition in position space as well by Fourier transforming, and we will use it in the main text. Now let us look at some expressions that will be useful in the computation of the entanglement entropy. Consider the expression
\bea
\begin{aligned}
 \left\langle\psi^{*}_{j}(\mathbf{X})\psi_{i}(\mathbf{X}^{\prime}) \right\rangle_{\mathrm{EE}} & = \mathcal{N}_{i} \mathcal{N}_{j} \int d^{3 N}P \  d^{3 N} P^{\prime} \   \left\langle A_{j}(\mathbf{-P}) \ A_{i}(\mathbf{P^{\prime}}) \right\rangle_{\mathrm{EE}}  \ \delta\left(\mathbf{P^{\prime}}^{2}-2 m U_{i}\right)\\
& \hspace{4cm} \delta\left(\mathbf{P}^{2}-2 m U_{j}\right)  \ \exp (\frac{i}{\hbar} \left(\mathbf{P^{\prime}} \cdot \mathbf{X^{\prime}-\mathbf{P} \cdot \mathbf{X}}\right) )
\end{aligned}
\eea
Using \eqref{atwopoint}, we get
\bea
\begin{aligned}
 \left\langle\psi^{*}_{j}(\mathbf{X})\psi_{i}(\mathbf{X}^{\prime}) \right\rangle_{\mathrm{EE}} & =\delta_{ij} \mathcal{N}_{i} \mathcal{N}_{j} \int d^{3 N}P \  d^{3 N} P^{\prime} \   \frac{\delta^{3 N}\left(\mathbf{P}-\mathbf{P}^{\prime}\right)}{ \delta\left(\mathbf{P}^{2}-\mathbf{P}^{\prime 2}\right)} \ \delta\left(\mathbf{P^{\prime}}^{2}-2 m U_{i}\right)\\
& \hspace{4cm} \delta\left(\mathbf{P}^{2}-2 m U_{j}\right)  \ \exp (\frac{i}{\hbar} \left(\mathbf{P^{\prime}} \cdot \mathbf{X^{\prime}-\mathbf{P} \cdot \mathbf{X}}\right) )
\end{aligned}
\eea
Let us focus on the Dirac delta part of the expression. We have
\bea
\begin{aligned}
\delta_{ij}  \frac{\delta^{3 N}\left(\mathbf{P}-\mathbf{P}^{\prime}\right)}{ \delta\left(\mathbf{P}^{2}-\mathbf{P}^{\prime 2}\right)} \  \delta\left(\mathbf{P^{\prime}}^{2}-2 m U_{i}\right)&\delta\left(\mathbf{P}^{2}-2 m U_{j}\right) \hspace{6cm}\\
&=\  \delta_{ij}  \frac{\delta^{3 N}\left(\mathbf{P}-\mathbf{P}^{\prime}\right)}{ \delta\left(\mathbf{P}^{2}-\mathbf{P}^{\prime 2}\right)} \ \delta\left(\mathbf{P^{\prime}}^{2}-2 m U_{i}\right)\delta\left(\mathbf{P}^{2}-2 m U_{i}\right) \\
&=\ \delta_{ij}  \frac{\delta^{3 N}\left(\mathbf{P}-\mathbf{P}^{\prime}\right)}{ \delta\left(\mathbf{P}^{2}-\mathbf{P}^{\prime 2}\right)} \ \delta\left(\mathbf{P}^{2}-\mathbf{P}^{\prime 2}\right)\delta\left(\mathbf{P^{\prime}}^{2}-2 m U_{i}\right)\\
&= \ \delta_{ij}  \delta^{3 N}\left(\mathbf{P}-\mathbf{P}^{\prime}\right) \delta\left(\mathbf{P^{\prime}}^{2}-2 m U_{i}\right)
\end{aligned}
\eea
Therefore, 
\bea
\begin{aligned}
 \left\langle\psi^{*}_{j}(\mathbf{X})\psi_{i}(\mathbf{X}^{\prime}) \right\rangle_{\mathrm{EE}} & =\delta_{ij} \mathcal{N}_{i} \mathcal{N}_{j} \int d^{3 N}P \  d^{3 N} P^{\prime} \  \delta^{3 N}\left(\mathbf{P}-\mathbf{P}^{\prime}\right) \  \delta\left(\mathbf{P^{\prime}}^{2}-2 m U_{i}\right) \\
&\hspace{6cm}  \exp (\frac{i}{\hbar} \left(\mathbf{P^{\prime}} \cdot \mathbf{X^{\prime}-\mathbf{P} \cdot \mathbf{X}}\right)) \\
&=\delta_{ij} \mathcal{N}^{2}_{i} \ \int d^{3 N}P \ \exp (\frac{i}{\hbar}  \left(\mathbf{P} \cdot (\mathbf{X^{\prime}- \mathbf{X}}\right))\ \delta\left(\mathbf{P}^{2}-2 m U_{j}\right) 
\end{aligned}
\eea
In particular, when $\mathbf{X^{\prime}= \mathbf{X}}$, we have
\bea
\begin{aligned}
 \left\langle\psi^{*}_{j}(\mathbf{X})\psi_{i}(\mathbf{X}) \right\rangle_{\mathrm{EE}} &= \delta_{ij} \mathcal{N}^{2}_{i} \ \int d^{3 N}P  \ \delta\left(\mathbf{P}^{2}-2 m U_{j}\right) \\
&=\delta_{ij}\  \mathcal{N}^{2}_{i} \ \frac{(2 \pi m U_{j} )^{3N / 2}}{\Gamma(3N/ 2) (2 m U_{j})}
\end{aligned}
\eea
Let us normalize this quantity by demanding that 
\bea
\int_{\mathcal{D}}  \left\langle\psi^{*}_{j}(\mathbf{X})\psi_{i}(\mathbf{X}) \right\rangle_{\mathrm{EE}}  \, d^{3 N}X = \delta_{ij}  \label{eqz}
\eea
This gives us the normalization condition
\bea
\mathcal{N}^{-2}_{j} = L^{3N} \ \frac{(2 \pi m U_{j} )^{3N / 2}}{\Gamma(3N/ 2) (2 m U_{j})} 
\eea
Now let us look at another important expression:
\bea
\begin{aligned}
&\int_{\mathcal{D}}  \,d^{3 N}X \,d^{3 N}X^{\prime}\left\langle \psi^{*}_{i}(\mathbf{X}) \psi_{j}(\mathbf{X}^{\prime}) \right\rangle_{\mathrm{EE}}\left\langle \psi_{k}(\mathbf{X}) \psi^{*}_{l}(\mathbf{X}^{\prime})\right\rangle_{\mathrm{EE}} \\
&=  \delta_{ij} \ \delta_{kl} \ \mathcal{N}^2_{i} \mathcal{N}^2_{k} \int_{\mathcal{D}}  \,d^{3 N}X \,d^{3 N}X^{\prime}\int d^{3 N}P \  d^{3 N} P^{\prime} \  \delta\left(\mathbf{P}^{2}-2 m U_{j}\right) \delta\left(\mathbf{P^{\prime}}^{2}-2 m U_{k}\right)\\
&\hspace{6.5cm}   \exp (\frac{i}{\hbar}  \left(\mathbf{P^{\prime}} \cdot (\mathbf{X- \mathbf{X}^{\prime}}\right)) \exp (\frac{i}{\hbar}  \left(\mathbf{P} \cdot (\mathbf{X^{\prime}- \mathbf{X}}\right)) \\
&=   \delta_{ij} \ \delta_{kl} \ \mathcal{N}^2_{i} \mathcal{N}^2_{k} \int_{\mathcal{D}}  \,d^{3 N}X \,d^{3 N}X^{\prime}\int d^{3 N}P \  d^{3 N} P^{\prime} \  \delta\left(\mathbf{P}^{2}-2 m U_{j}\right) \delta\left(\mathbf{P^{\prime}}^{2}-2 m U_{k}\right)\\
&\hspace{6.5cm}   \exp (\frac{i}{\hbar}  \left(\mathbf{X^{\prime}} \cdot (\mathbf{P- \mathbf{P}^{\prime}}\right)) \exp (\frac{i}{\hbar}  \left(\mathbf{X} \cdot (\mathbf{P^{\prime}- \mathbf{P}}\right)) \\
&=   \delta_{ij} \ \delta_{kl} \ h^{6 N} \ \mathcal{N}^2_{i} \mathcal{N}^2_{k}\int d^{3 N}P \  d^{3 N} P^{\prime} \  \delta\left(\mathbf{P}^{2}-2 m U_{j}\right) \delta\left(\mathbf{P^{\prime}}^{2}-2 m U_{k}\right)  \left[\delta_{\mathcal{D}}^{3 N}(\mathbf{P-P^{\prime}})\right]^2 \label{z2eq1}
\end{aligned}
\eea
where we have defined
\bea
\delta_{\mathcal{D}}^{3 N}(\mathbf{K}) \equiv h^{-3 N} \int_{\mathcal{D}} d^{3 N} X \exp (i \mathbf{K} \cdot \mathbf{X} / \hbar) \label{deltaD}
\eea
Let us look at the integral in the last line of \eqref{z2eq1}. It is convenient to define this integral as a separate quantity as follows:
\bea
\Phi_{ij} \equiv \int d^{3 N}P \  d^{3 N} P^{\prime} \  \delta\left(\mathbf{P}^{2}-2 m U_{i}\right) \delta\left(\mathbf{P^{\prime}}^{2}-2 m U_{j}\right)  \left[\delta_{\mathcal{D}}^{3 N}(\mathbf{P-P^{\prime}})\right]^2
\eea
Now let us first look at the case where $i=j$. To do the integral explicitly, we will work in the low density limit where $Na^3 \ll L^3$. The factor $\delta_{\mathcal{D}}^{3 N}(\mathbf{P-P^{\prime}})$ will effectively act as a Dirac delta function throughout the range of variable $\mathbf{P-P^{\prime}}$. Therefore, it suffices to use the replacement \cite{Srednicki1}
\bea
\left[\delta_{\mathcal{D}}^{3 N}(\mathbf{P})\right]^{2} \rightarrow(L / h)^{3 N} \delta^{3 N}(\mathbf{P}) \label{eqdeltaapprox}
\eea
this gives us
\bea
\begin{aligned}
\Phi_{ij} &=  \int d^{3 N}P \  d^{3 N} P^{\prime} \  \delta\left(\mathbf{P}^{2}-2 m U_{i}\right) \delta\left(\mathbf{P^{\prime}}^{2}-2 m U_{i}\right)  \delta^{3 N}(\mathbf{P-P^{\prime}}) \\
&=  \int d^{3 N}P \  \delta\left(\mathbf{P}^{2}-2 m U_{i}\right) \delta\left(\mathbf{P}^{2}-2 m U_{i}\right) \\
&=  \int d^{3 N}P \  \delta\left(\mathbf{P}^{2}-2 m U_{i}\right) \\
&=  \frac{(2 \pi m U_{i} )^{3N / 2}}{\Gamma(3N/ 2) (2 m U_{i})} \label{z2eq2}
\end{aligned}
\eea
Now let us look at the case where $i \neq j$. If we used the same replacement as in \eqref{eqdeltaapprox}, then we will see that the integral will be non-zero only when $U_i=U_j$. However, we can see from the explicit form\footnote{Note that the integration range that defines \eqref{deltaD} is non-trivial, which is what makes it different from an ordinary delta function. The explicit form, as can be checked by doing the integral explicitly for a 1-D case, is slightly delocalized.} of $\delta_{\mathcal{D}}^{3 N}(\mathbf{P-P^{\prime}})$ that the integral \eqref{z2eq1} will be non-zero when $U_i$ and $U_j$ are sufficiently close to each other. In fact, it turns out that these states are the ones that dominate the entanglement entropy calculation. Therefore, we will have to use a more precise replacement to do the computation. It suffices to use a Gaussian approximation as follows (see \cite{Srednicki1} for a closely related but distinct expression)
\bea
\delta_{\mathcal{D}}^{3 N}\left(\mathbf{P}-\mathbf{P}^{\prime}\right) \simeq(L / h)^{3 N} \exp \left[-\left(\mathbf{P}-\mathbf{P}^{\prime}\right)^{2} L^{2} / 4 \pi \hbar^{2}\right]\label{eq3ddelta}
\eea
Using the Gaussian approximation, we can see from \cite{Srednicki1} that  
\bea
\begin{aligned}
\Phi_{ij} &\simeq \Phi_{ii}\exp \left[-m\left(U_{i}-U_{j}\right)^{2} L^{2} / 8 \pi \hbar^{2} U_{i}\right]
\end{aligned}
\eea
Substituting this into \eqref{z2eq1} we get 
\bea
\begin{aligned}
&\int_{\mathcal{D}}  \,d^{3 N}X \,d^{3 N}X^{\prime}\left\langle \psi^{*}_{i}(\mathbf{X}) \psi_{j}(\mathbf{X}^{\prime}) \right\rangle_{\mathrm{EE}}\left\langle \psi_{k}(\mathbf{X}) \psi^{*}_{l}(\mathbf{X}^{\prime})\right\rangle_{\mathrm{EE}}  \\
 &=   \delta_{ij} \ \delta_{kl} \ (Lh)^{3 N} \ \mathcal{N}^2_{i} \mathcal{N}^2_{k} \  \frac{(2 \pi m U_{k} )^{3N / 2}}{\Gamma(3N/ 2) (2 m U_{k})}\ \exp \left[-m\left(U_{j}-U_{k}\right)^{2} L^{2} / 8 \pi \hbar^{2} U_{j}\right] \\
&=   \delta_{ij} \ \delta_{kl} \left(\frac{L}{h}\right)^{-3N}\frac{\Gamma(3N/ 2) (2 m U_i)}{(2 \pi m U_i )^{3N / 2}} \exp \left[-m\left(U_{j}-U_{k}\right)^{2} L^{2} / 8 \pi \hbar^{2} U_{j}\right] \\
&=  \delta_{ij} \ \delta_{kl} \ Z_{i} \exp \left[-m\left(U_{j}-U_{k}\right)^{2} L^{2} / 8 \pi \hbar^{2} U_{j}\right] \label{eqZcondition}
\end{aligned}
\eea
where we have defined
\bea
Z_{i} = \left(\frac{L}{h}\right)^{-3N}\frac{\Gamma(3N/ 2) (2 m U_i)}{(2 \pi m U_i )^{3N / 2}}
\eea
When $|U_j-U_k|/U_j$ smaller than or equal to $\left(\hbar^{2} / m U_{j} L^{2}\right)^{1 / 2}$, the exponential can effectively be replaced by 1 and we will get
\bea
\begin{aligned}
&\int_{\mathcal{D}}  \,d^{3 N}X \,d^{3 N}X^{\prime}\left\langle \psi^{*}_{i}(\mathbf{X}) \psi_{j}(\mathbf{X}^{\prime}) \right\rangle_{\mathrm{EE}}\left\langle \psi_{k}(\mathbf{X}) \psi^{*}_{l}(\mathbf{X}^{\prime})\right\rangle_{\mathrm{EE}} =  \delta_{ij} \ \delta_{kl} \ Z_{i} 
\end{aligned}
\eea
When the eigenstates are sufficiently close to each other in the energy spectrum, we can also see that
\bea
\begin{aligned}
&\int_{\mathcal{D}}  \ \left\langle \psi^{*}_{i^{}_1}(\mathbf{X}_1) \psi_{i^{}_2}(\mathbf{X}_2) \right\rangle_{\mathrm{EE}}\left\langle \psi_{i^{}_3}(\mathbf{X}_2) \psi^{*}_{i^{}_4}(\mathbf{X}_3)\right\rangle_{\mathrm{EE}} \cdots \left\langle \psi_{i^{}_{2n-1}}(\mathbf{X}_{n}) \psi^{*}_{i^{}_{2n}}(\mathbf{X}_1)\right\rangle_{\mathrm{EE}}\prod_{j=1}^{n}\,d^{3 N}X_j \\
&\simeq \left(\prod_{j=1}^{n}  \delta_{i^{}_{2j-1}i^{}_{2j}} \right)\left(\prod_{j=1}^{n-1}  \ Z_{i^{}_{2j-1}} \right) \label{eqZcondition1}
\end{aligned}
\eea
\section{Sub-leading corrections to the entanglement entropy}
\subsection{Corrections to \eqref{traceeq2}}
\label{appcorrectdelta}
Let us look at the corrections to \eqref{eqpurityfinal00} when we relax the conditions ${\Delta}_L \lesssim \Delta^{\prime}_{L}$ and ${\Delta}_S \lesssim \Delta^{\prime}_{S}$. We will start with the first sum on the RHS of \eqref{traceeq000}. When ${\Delta}_L \gtrsim \Delta^{\prime}_{L}$, the dominant contribution to the sum will be given by
\bea
\sum_{i^{}_{1_{S}},i^{}_{2_{S}}} \sum_{\tilde{i}^{}_{1_{L}},\tilde{i}^{}_{2_{L}}} \left| c^{}_{ i^{}_{1_{S}}\tilde{i}^{}_{1_{L}}}\right|^2 \left| c^{}_{ i^{}_{2_{S}}\tilde{i}^{}_{2_{L}}}\right|^2  Z_{\tilde{i}^{}_{1_L}}
\eea
Here, the indices $\tilde{i}^{}_{1_{L}}$ and $\tilde{i}^{}_{2_{L}}$ are restricted to the set that satisfy the condition 
\bea
\left|U_{\tilde{i}^{}_{1_{L}}}-U_{\tilde{i}^{}_{2_{L}}}\right|  \lesssim U_{\tilde{i}^{}_{1_{L}}} \left(\hbar^{2} / m U_{\tilde{i}^{}_{1_{L}}} {L^{\prime}}^{2}\right)^{1 / 2}
\eea
Therefore, the first term on the RHS of \eqref{eqpurityfinal00} will turn out to be
\bea
 \frac{1 }{\left(\sum_{i^{}_{1_{S}},i^{}_{1_{L}}} \left| c^{}_{ i^{}_{1_{S}}i^{}_{1_{L}}}\right|^2 \right)^2}\sum_{i^{}_{1_{S}},i^{}_{2_{S}}} \sum_{\tilde{i}^{}_{1_{L}},\tilde{i}^{}_{2_{L}}} \left| c^{}_{ i^{}_{1_{S}}\tilde{i}^{}_{1_{L}}}\right|^2 \left| c^{}_{ i^{}_{2_{S}}\tilde{i}^{}_{2_{L}}}\right|^2  Z_{\tilde{i}^{}_{1_L}}
\eea
We can crudely approximate the coefficients as in \cite{Srednicki1}:
\bea
 c^{}_{ i^{}_{1_{S}}i^{}_{1_{L}}} \sim \frac{1}{\sqrt{\bar{n}_L \Delta_L}} \frac{1}{\sqrt{\bar{n}_S \Delta_S}}
\eea
where $\bar{n}_L$ and $\bar{n}_S$ are the energy level densities near $\bar{U}_L$ and $\bar{U}_S$ respectively. This will give us
\bea
\begin{aligned}
 \frac{\sum_{\tilde{i}^{}_{1_{L}},\tilde{i}^{}_{2_{L}}}  Z_{\tilde{i}^{}_{1_L}}}{\sum_{i^{}_{1_{L}},i^{}_{2_{L}}}  } &\simeq  \frac{\sum_{\tilde{i}^{}_{1_{L}}}  Z_{\tilde{i}^{}_{1_L}}\left(n_{\tilde{i}^{}_{1_L}} U_{\tilde{i}^{}_{1_{L}}}\left(\hbar^{2} / m U_{\tilde{i}^{}_{1_{L}}} {L^{\prime}}^{2}\right)^{1 / 2}\right)}{\left(\bar{n}_L \Delta_L\right)^2  }
\end{aligned}\label{eqappendix00}
\eea
Where  we have used $n_{\tilde{i}^{}_{1_L}}$ to denote the energy level density around $U_{\tilde{i}^{}_{1_L}}$. Using the explicit forms of $ Z_{i^{}_{1_L}}$ and the energy level density \cite{Srednicki1},
\bea
n_{i}=\frac{1}{\Gamma(3 N / 2) U_{i}}\left(\frac{m L^{2} U_{i}}{2 \pi \hbar^{2}}\right)^{3 N / 2}, \label{eqleveldensity}
\eea
we can simplify \eqref{eqappendix00} to get
\bea
\begin{aligned}
\left(\frac{2m\left(\hbar^{2} / m  {L^{\prime}}^{2}\right)^{1 / 2}}{\left(2\pi\right)^{3N_L}}\right)&\left(\frac{\sum_{i^{}_{1_{L}}}  U_{\tilde{i}^{}_{1_{L}}}^{1/2} }{  \left(\bar{n}_L \Delta_L\right)^2}  \right) \\
& = \left(\frac{2m\left(\hbar^{2} / m  {L^{\prime}}^{2}\right)^{1 / 2}}{\left(2\pi\right)^{3N_L}}\right)\left[\frac{\sum_{i^{}_{1_{L}}} \left(\bar{U}_{L} + \left(U_{\tilde{i}^{}_{1_{L}}}-\bar{U}_{L}\right)\right)^{1/2}  }{  \left(\bar{n}_L \Delta_L\right)^2}  \right] \\
& = \left(\frac{2m\left(\hbar^{2} / m  {L^{\prime}}^{2}\right)^{1 / 2}}{\left(2\pi\right)^{3N_L}}\right)\left(\frac{ \bar{U}_{L}^{1/2}  }{  \left(\bar{n}_L \Delta_L\right)^2}  \right)\left[\sum_{i^{}_{1_{L}}}\left(1 + O\left(\frac{\Delta_L}{\bar{U}_{L}}\right)\right)^{\frac{1}{2}}\right]\\
& = \left(\frac{2m\left(\hbar^{2} / m  {L^{\prime}}^{2}\right)^{1 / 2}}{\left(2\pi\right)^{3N_L}}\right)\left(\frac{ \bar{U}_{L}^{1/2}  }{  \left(\bar{n}_L \Delta_L\right)}  \right)\left(1 + O\left(\frac{\Delta_L}{\bar{U}_{L}}\right)\right)^{\frac{1}{2}}
\end{aligned}\label{eqappendix01}
\eea
As $\frac{\Delta_L}{\bar{U}_{L}} \ll 1$, we can drop the last factor. Substituting the explicit form of $\bar{n}_L$ (from \eqref{eqleveldensity}) and simplifying, we get the following expression
\bea
\left(\frac{L^{\prime}}{h}\right)^{-3N_L}\frac{\Gamma(3N_L/ 2) (2 m \bar{U}_L)}{(2 \pi m \bar{U}_L)^{3N_L / 2}} \left(\frac{\Delta^{\prime}_L }{\Delta_L}\right)
\eea
Comparing this with \eqref{eqpurityfinal00} and \eqref{eqaveng1}, we can see that $\text{Tr}(\tilde{\rho}_{L}\mathcal{I}_{L}) $ has picked up an additional factor of $\frac{\Delta^{\prime}_L }{\Delta_L}$. Now let us look at the corrections to the entanglement entropy at early times. From \eqref{entropytranseq}, we have
\bea
\begin{aligned}
\left \langle S (\rho_{L}) \right \rangle_{\mathrm{EE}} = -\log{\text{Tr}(\tilde{\rho}_{L}\mathcal{I}_{L}) }
\end{aligned}
\eea
Therefore
\bea
\begin{aligned}
\left \langle S (\rho_{L}) \right \rangle_{\mathrm{EE}} &\simeq N_L\left\{\log{\left[V_L\left(\frac{4\pi m\bar{U}_{L}}{3 h^2 N_L}\right)^{3/2}\right]} +\frac{3}{2}\right\} + O\left(\log{\frac{\Delta^{\prime}_L }{\Delta_L}}\right)\\
& = N_L\log{\left[V_L\left(\frac{4\pi m\bar{U}_{L}}{3 h^2 N_L}\right)^{3/2}\right]}\left(1+O\left(\frac{1}{N_L}\right)\right) +\frac{3N_L}{2} +O\left(\frac{\log{L^{\prime}}}{N_L}\right)\\
& \hspace{7cm}+O\left(\frac{\log{N_L}}{N_L}\right)  +O\left(\log{\frac{\Delta_L }{\bar{U}_L}}\right) 
\end{aligned}
\eea 
We have used the explicit form of \eqref{eqemergedelta1} to get to the second equality. We can see that the corrections can make significant contribution to \eqref{finalentropyeq1} only when
\bea
\log{\frac{\Delta_L }{\bar{U}_L}} \sim N_L
\eea
When ${\Delta}_L \gtrsim \Delta^{\prime}_{L}$, the only possibility is to have $\frac{\Delta_L }{\bar{U}} \sim \exp{N_L}$. As $\frac{\Delta_L }{\bar{U}_L} \ll 1$, this condition will never be satisfied. Therefore, the corrections to the entanglement entropy will always be sub-leading in $N_L$. We can perform a similar analysis on the entanglement entropy at late times and show that the corrections are sub-leading as well.

\subsection{Corrections to \eqref{eqaveng1} and \eqref{eqaveng2}}
\label{correctionappendix}
In this appendix, we will explicitly calculate the quantities on the RHS of \eqref{eqpurityfinal00}. Let us begin with the following expression:
\bea
\text{Tr}(\tilde{\rho}_{L}\mathcal{I}_{L})  = \frac{\sum_{i^{}_{1_{S}},i^{}_{1_{L}}} \left| c^{}_{ i^{}_{1_{S}}i^{}_{1_{L}}}\right|^2 Z_{i^{}_{1_L}}}{\sum_{i^{}_{1_{S}},i^{}_{1_{L}}} \left| c^{}_{ i^{}_{1_{S}}i^{}_{1_{L}}}\right|^2 } \label{appendixeq1}
\eea
where
\bea
Z_{i^{}_{L}} = \left(\frac{L^{\prime}}{h}\right)^{-3N_{L}}\frac{\Gamma(3N_{L}/ 2) (2 m U_{i^{}_{L}})}{(2 \pi m U_{i^{}_{L}})^{3N_{L} / 2}}  \label{appendixeq4001}
\eea
Let us focus on the following expression:
\bea
\begin{aligned}
(2 m U_{i^{}_{L}})^{\frac{3N_{L}-1}{2}} &= \left(2 \pi m \left(\bar{U}_{L} + \left(U_{i^{}_{L}}-\bar{U}_{L}\right)\right) \right)^{\frac{3N_{L}-1}{2}} \\
& =\left(2 \pi m \bar{U}_{L}\right)^{\frac{3N_{L}-1}{2}}\left(1 + \frac{\left(U_{i^{}_{L}}-\bar{U_{L}}\right)}{\bar{U}_{L}}\right)^{\frac{3N_{L}-1}{2}}
\end{aligned} \label{appendixeq4}
\eea
Therefore, we can rewrite \eqref{appendixeq4001} as
\bea
 Z_{i^{}_{L}} \simeq \left(\frac{L^{\prime}}{h}\right)^{-3N_{L}}\frac{\Gamma(3N_{L}/ 2) (2 m \bar{U}_{L})}{(2 \pi m \bar{U}_{L})^{3N_{L} / 2}}\left(1 +O\left( \frac{\Delta_L}{\bar{U}_{L}}\right)\right)^{\frac{3N_{L}-1}{2}} \label{appendixeq2}
\eea
Using \eqref{appendixeq1} and \eqref{appendixeq2}, we get
\bea
\text{Tr}(\tilde{\rho}_{L}\mathcal{I}_{L}) \simeq \left(\frac{L^{\prime}}{h}\right)^{-3N_{L}}\frac{\Gamma(3N_{L}/ 2) (2 m \bar{U}_{L})}{(2 \pi m \bar{U}_{L})^{3N_{L} / 2}}\left(1 + O\left(\frac{\Delta_L}{\bar{U}_{L}}\right)\right)^{\frac{3N_{L}-1}{2}} \label{appendixeq3}
\eea
As we cannot make any claims about $\frac{N_L\Delta_L}{\bar{U}_{L}}$, we cannot drop the last factor. However, it turns out that the contribution of this factor to the \textit{entanglement entropy} is sub-leading in $N_L$. To see this, let us recall that the entanglement entropy at early times is given by \eqref{entropytranseq}:
\bea
\begin{aligned}
\left \langle S (\rho_{L}) \right \rangle_{\mathrm{EE}} &= \lim_{n \to  1} \frac{1}{1-n}\log\left[\frac{\text{Tr} \left\langle \rho^{n}_{L} \right\rangle_{\mathrm{EE}}}{\left(\text{Tr} \left\langle \rho_{L} \right\rangle_{\mathrm{EE}}\right)^n}\right] \\
&= -\log{\text{Tr}(\tilde{\rho}_{L}\mathcal{I}_{L}) }
\end{aligned}
\eea
Therefore, by taking a log of \eqref{appendixeq3} and rearranging the terms, we get
\bea
 \left \langle S (\rho_{L}) \right \rangle_{\mathrm{EE}} \simeq - \log{\left[ \left(\frac{L^{\prime}}{h}\right)^{-3N_L}\frac{\Gamma(3N_L/ 2) (2 m \bar{U}_L )}{(2 \pi m \bar{U}_{L})^{3N_L / 2}}\right]} - \log{\left(1 + O\left(\frac{\Delta_L}{\bar{U}_{L}}\right)\right)^{\frac{3N_L-1}{2}}}
\eea
For large $N_L$ and $\frac{\Delta_L}{\bar{U}_{L}} \ll 1$, we can simplify this expression to get
\bea
\left \langle S (\rho_{L}) \right \rangle_{\mathrm{EE}} \simeq  N_L\log{\left[V_L\left(\frac{4\pi m\bar{U}_{L}}{3 h^2 N_L}\right)^{3/2}\right]} +\frac{3 N_L}{2}\left(1 +O\left(\frac{\Delta_L}{\bar{U}_{L}}\right)\right) + O\left(\frac{\log{N_L}}{N_L}\right)
\eea
As $\frac{\Delta_L}{\bar{U}_{L}} \ll 1$, we can see that the leading part of the above expression matches precisely with \eqref{finalentropyeq1}. Repeating the same calculations for the smaller box, we can see that entanglement entropy at late times will also have sub-leading corrections to \eqref{finalentropyeq1} in $\frac{\Delta_S}{\bar{U}_{S}}$. Therefore, for all our purposes, it is legitimate to drop the last factor in \eqref{appendixeq3} and this gives the expressions \eqref{eqaveng1} and \eqref{eqaveng2}.

\end{document}